Title page

Tita Alissa Bach[a*1], Aleksandar Babic [a*], Narae Park [a*2], Tor Sporsem[d], Rasmus Ulfsnes[d], Henrik Smith-Meyer[b], Torkel Skeie[c]

[a]Group Research and Development, DNV, Høvik, Norway

[b]Maritime Production Operations, DNV, Høvik, Norway

[c]Technical Support Norway, DNV, Høvik, Norway

[d]Department of Software Engineering, Safety and Security, Sintef Digital, Trondheim, Norway

[1]Corresponding Author: tita.alissa.bach@dnv.com

*These authors have contributed equally to this work

Word count: 15507

[2]Current affiliation: House of Math, Oslo, Norway

**Title: Using LLM-Generated Draft Replies to Support Human Experts in Responding to Stakeholder Inquiries in Maritime Industry: A Real-World Case Study of Industrial AI**

**Abstract** (188 words):

This study investigated the utility of Large Language Models (LLMs) in supporting human experts in the maritime industry by generating draft replies to stakeholder inquiries. Using a mixed-methods approach, observations, interviews, a survey, and text similarity analysis, we examined how human experts perceive LLM drafts, how drafts align with final replies, and their potential impact on workflows and stakeholder trust. Findings show that while LLM drafts can improve efficiency, linguistic consistency, and time management, they frequently require manual adjustments to ensure accuracy and relevance. Case handlers' attitudes were generally "skeptical but curious," reflecting cautious adoption and emphasizing the role of managerial support in fostering trust and engagement. Text similarity analyses indicated moderate alignment between drafts and final replies, reinforcing the need for human oversight. Overall, LLMs are not yet suitable for independent use in safety-critical settings but can serve as valuable augmentative tools. By leveraging human expertise alongside LLM capabilities, organizations can enhance workflow efficiency while maintaining quality, accuracy, and stakeholder trust. These findings provide actionable guidance for implementing AI responsibly in specialized industrial contexts and for designing human-AI collaboration that balances automation benefits with critical human judgment.

**A shortened version of the title for a running head**: Supporting Maritime Experts with LLM Draft Replies

# 1. Introduction

The maritime industry is a significant contributor to worldwide trade, transportation, and the global economy. It is a complex and dynamic sector characterized by multifaceted operations, increasingly more stringent regulations, and a diverse array of stakeholders that include shipowners, port authorities, regulatory bodies, suppliers, and ship management organizations (Akpinar and Ozer-Caylan 2022). In such a complex environment, effective communication is paramount, requiring timely, accurate, and contextually appropriate responses to stakeholder inquiries. This can often be a daunting task for human experts due to the volume, complexity, time-critical, and especially nicheness of the information involved. Artificial intelligence (AI) has the potential to alleviate these communication challenges in the maritime industry by assisting human experts with the demands of stakeholder interactions.

In recent years, advancements in Industrial AI-enabled systems have introduced promising tools to support and enhance human capabilities in various professional domains. Here, we define AI-enabled systems as any system that contains or relies on one or more AI components, distinct units of software that perform a specific function or task within an AI-enabled system and consist of a set of AI models, data, and algorithms (Tita Alissa Bach, Kaarstad, et al. 2024; DNV 2023). We use "Industrial AI" here to refer to the specialized application of AI-enabled systems within specific industries – such as energy and maritime – to optimize processes, enhance decision-making, and drive innovation. Unlike general AI applications such as customer AI, which often prioritizes broad adaptability and versatility across various domains, Industrial AI is tailored to address the unique challenges and requirements of a particular industry (DNV 2024).

Industrial AI must be deployable with acceptable risk to be robust and trustworthy. This requires an understanding of both conventional industry risks and new AI-specific risks. Key characteristics of Industrial AI include: the AI's technical performance; its role and agency within the system; its interaction with other components; and its technical, legal, and ethical impact on stakeholders. A thorough understanding of these risks ensures that Industrial AI systems align with industry-specific standards and deliver value while maintaining safety, reliability, and trust.

One advancement in Industrial AI-enabled systems that is pertinent to the maritime industry is the development of Large Language Models (LLMs), a type of Generative AI. These models, such as OpenAI's GPT-4, can generate human-like text based on the prompts they receive. They have demonstrated potential in various applications, including drafting emails, creating content, and even coding (Thiergart, Huber, and Übellacker 2021; Chiarello et al. 2024; Urlana et al. 2024; C. Jiang, Wang, Ma, et al. 2024; J. Jiang, Wang, Shen, et al. 2024). This could potentially address many of the challenges to effective communication in the maritime industry.

However, LLMs' efficacy and usefulness in highly specialized, safety-critical fields, such as the maritime industry, remain area for exploration (Hodne et al. 2024; Reddy, Sinha, and Park 2024). While LLMs have shown great potential in various industries such as healthcare, law, finance, and education (Urlana et al. 2024; Meng et al. 2024; Lai et al. 2024; Nie et al. 2024; S. Wang et al. 2024), their reported weaknesses – such as factual inaccuracies (C. Wang et al. 2023), hallucinations (Huang et al. 2025; Dahl et al. 2024), and a lack of contextual understanding (Abdurahman et al. 2024; Chen et al. 2024), pose significant risks ("Beyond words? The possibilities, limitation, and risks of large language models" ; Lappin 2024). In the maritime industry, where precision and adherence to regulations are critical, the consequences of LLM inaccuracies could be severe. Erroneous information in compliance documentation, misinterpretation of international maritime laws, or incorrect advice on safety procedures could lead to regulatory penalties, operational inefficiencies, or even accidents that threaten both human lives and the environment. These weaknesses highlight the need for robust validation processes and human oversight when deploying LLMs in these high-stakes contexts.

Fortunately, LLMs need not to be perfectly accurate to be valuable. They can, for example, serve as tools for generating drafts, offering preliminary suggestions, or enhancing communication by translating technical information into a more accessible language. As long as LLMs transmit the appropriate information, they can provide value.

Our real-world case study investigates the integration of Large Language Models (LLMs) into the workflow of maritime technical experts (i.e., case handlers) at DNV Maritime's Technical Help Desk (THD). We focus on the role and impact of LLM-generated draft replies in assisting case handlers when responding to stakeholder inquiries. To achieve this, our study aims to comprehensively evaluate the utility of LLM-generated draft replies through two complementary perspectives: the subjective experiences of human experts and the objective textual similarity between LLM drafts and case handlers' final responses. These dual objectives lead to the following two research questions, each investigated through distinct methodological approaches.

*RQ 1: How do case handlers perceive the usefulness of LLM drafts, and how can these drafts support their workflows?*

To address the first question, we adopt a mixed-methods approach combining qualitative preliminary research (observations and interviews) with quantitative survey data from case handlers. This question emphasizes the subjective experiences and insights of case handlers, capturing both the perceived usefulness and the various ways LLM drafts may assist their tasks (e.g., summarizing inquiries, standardizing replies, or providing reference points).

*RQ2: How similar are the texts generated by the LLM to the final replies sent by case handlers, and what does this similarity imply about the drafts' potential utility?*

To address the second question, we employ a computational approach that quantitatively assesses the alignment between LLM drafts and case handlers' final responses using text similarity metrics. The degree of similarity between LLM drafts and case handlers' final responses can serve as a proxy for the drafts' practical utility. Higher textual similarity may suggest that the drafts were useful as efficient tools requiring minimal modifications and aligning well with case handlers' intentions. However, recognizing that similarity may not necessarily correspond to utility (e.g., cases where LLM drafts are useful as a starting point even when sent replies differ significantly), we broadly explore whether textual similarity between LLM drafts and sent replies can serve as a proxy for the drafts' practical relevance.

We investigate how LLMs can augment human expertise and streamline workflows in the maritime industry as a case study for Industrial AI applications (Seo et al. 2024; Ulfsnes et al. 2024). We focus on their potential utility in drafting initial responses to inquiries, synthesizing complex information into accessible formats, and identifying patterns or insights from vast amounts of data. By evaluating these applications, we seek to understand how these draft replies (“LLM drafts”) can serve as a starting point rather than final deliveries, while remembering that the role of human judgment is essential for refining and validating the AI-generated content (Tita A Bach, Kristiansen, et al. 2024; Duenas and Ruiz 2024). Through this lens, we aim to uncover practical use cases that balance the strengths and weaknesses of LLMs and align their capabilities with real-world needs in maritime operations (Tita A Bach, Kristiansen, et al. 2024).

The findings from our case study are intended to provide insights into the potential benefits and limitations of using LLMs to support communication in a real-world specialized industry setting. We aim to offer practical recommendations and future work for integrating such technologies to enhance operational efficiency and stakeholder engagement for industrial use cases. Our case study also contributes to the broader discourse on human-AI collaboration by highlighting the transformative potential of AI-enabled systems in highly specialized domains (Tita A Bach, Kristiansen, et al. 2024).

# 2. Background

## 2.1. The case study: Direct Access to Technical Experts (DATE) system

We conducted our case study within DNV Maritime ("DNV Maritime"), a division of DNV that is responsible and specializes in classification, technical assurance, software, and advisory services for the maritime industry. DNV as an organization operates as an independent third party, providing services such as quality assurance and risk management in safety-critical industries and critical infrastructure. DNV Maritime plays a critical role in ensuring the safety, compliance, and technical integrity of ships and maritime operations worldwide. It provides classification, certification, and technical advisory services based on rigorous rules and international regulations.

DNV Maritime has a flagship service called The Technical Help Desk (THD) branded as "Direct Access to Technical Experts" (DATE system) ("DATE system"). The DATE system is a specialized service designed to assist with most technical inquiries related to Fleet in Service (FiS), Certification of Materials and Components (CMC), and Newbuilding projects. The goal is to provide timely and expert support to DNV Maritime's stakeholders (customers), ensuring their technical issues are resolved efficiently and effectively. In practice, the DATE system functions as a query-answering workflow: when a stakeholder (customer) submits a technical question, such as how to interpret a classification rule or whether a specific equipment modification complies with regulations, a case handler (a DNV technical expert) is assigned to respond.

In brief, the query-answering process on DATE system works as the following in practice:

- Submission of an inquiry. An inquiry from a stakeholder is submitted on the DATE system's digital platform.
- Case assignment. Each inquiry is logged as a case and assigned to a case handler, a DNV technical expert with relevant domain knowledge. Case handlers are located across global offices (e.g., Norway, Germany, Singapore, USA), enabling 24/7 coverage.
- Initial Assessment. The case handler reviews the query, assesses its scope, and determines whether they can respond independently based on, for example, relevant rules, procedures, and internal guidelines.
- Expert Collaboration (if needed). Due to the highly specialized nature of maritime engineering and regulation, spanning over 650 distinct fields of expertise, case handlers frequently consult internal subject-matter experts (SMEs). This collaboration ensures technical accuracy, especially for complex or novel cases. The SMEs may be engaged asynchronously via the DATE platform and discussions are documented within the case file for traceability.
- Response Drafting and Validation. Based on their own knowledge and input from SMEs, the case handler drafts a response. The response is reviewed if necessary by the SMEs before finalizing and sending it to the stakeholder through the DATE system. For Fleet in Service (FiS) cases as the most common type, DNV guarantees a response within 24 hours. The system handles approximately 60,000 FiS cases annually, with a 96% on-time response rate.

- Knowledge Retention. All interactions, decisions, and expert consultations are stored in the DATE platform. This creates an auditable trail and supports future case resolution through precedent.

Experts use the DATE system platform to ensure consistent and efficient handling of both internal and external inquiries. These inquiries are made primarily by the ship management organizations within the maritime industry (see Table 1 for examples of real-world inquiries and replies). Ship management organizations are companies that oversee and manage the operational, technical, and administrative aspects of ships on behalf of their owners, ensuring compliance with international and national regulations and efficient performance. Most of the inquiries come from ship management personnel. In some cases, the inquiries can also come from ship owners, captains, and technical personnel (in this case study, all are included in "stakeholders").

Table 1. Examples of real-world inquiries and sent replies related to radar malfunctions

| **Example inquiry** | **Example sent reply** |
|---|---|
| [ship name] has reported a defect X-band radar and is calling [a port] tomorrow. Repair is ordered and radar is expected to be operational upon departure. We expect your surveyor on board in order to close annual survey so expect they can close out radar issue. | Thank you for contacting DNV's DATE® service. The notification with regards the malfunctioning X-Band radar is well received.<br><br>It is understood that the repair has been scheduled upon arrival to next port, where our surveyor shall attend the vessel as well. Upon the repairs, the attending surveyor can close the finding. Hope the above clarifies. |
| Regret to inform that the monitor for the x-band radar is recently out of order. Reason unknown. We are in contact with the maker in order to arrange a new monitor. The maker is able to supply a new monitor, but same is available only on/abt [a specific date], agw[1].<br><br>The vessel is recently at anchorage [a port name]. Next [a port name] / ETA [a specific date]. | Thank you for contacting DNV's DATE® service. We understand that you are asking about a defective x-band radar. DNV may issue a Condition of Authority (CA) and relevant Short Term / Conditional Certificate, for vessels for which remote survey is possible. Prior Flag acceptance from [a flag name] is not required.<br><br>We propose to issue the following Condition of Authority (CA), in addition to ST/Conditional Certificate valid for one month, on behalf of the flag:<br>The malfunctioning Radar to be repaired within due date. In the meantime, the following interim measures to be adhered to:<br>Risk assessment to be carried out by the master and mitigating actions to be put in place. All other navigational equipment including the remaining radar to be in good working order. Suitable warning notices to be placed on the bridge and the unavailable radar. Additional lookout(s) to be posted when in congested waters or in reduced visibility. Entry of the defective radar shall be recorded in the ship's logbook. Failure of radar to be taken into account in Master's instructions to watchkeepers and the voyage |
| | planning process. Coastal/port state authorities and pilots to be advised of the situation prior to arrival and departure from port.<br><br>Please upload documentation showing compliance with the above-listed conditions, including a master statement and inform the expected rectification date.<br><br>Kindly note that a fee will be charged for the issuance of CA. If your company requires a PO number, please include this information before confirming the CA.<br><br>If received documentation is to DNV surveyor satisfaction, you will be notified by the system once the survey statement confirming the CA and ST/Conditional certificate issuance is uploaded. |
| [1]All going well (a maritime communication phrase, indicating that the statement is contingent on no unforeseen issues arising, providing a condition assurance of progress or completion) | |

The highly niche nature of the stakeholders' needs makes responding to these inquiries challenging. Table 1 provides illustrative examples of real maritime stakeholder communications, specifically the inquiries and sent replies related to radar malfunctions. We include this table to demonstrate the highly technical and domain-specific nature of inquiries and replies in this industry, which motivated our case study, methodological choices and the need for LLM support. Case handlers often have to consult with colleagues and other experts and search for similar past cases to guide their replies. Historical records of previous inquiries can be invaluable for providing context and ensuring accurate, consistent, and timely responses to these complex and specialized requests. Not all inquiries are highly complex or niche; some – such as requests for specific regulations that apply to a particular case – are more straightforward. Even in these simpler scenarios, referring to past cases can help ensure consistency and efficiency in responses. However, the sheer volume of accumulated cases makes it difficult, or at times impossible, to efficiently identify relevant precedents. The database contains over 600,000 past cases which poses significant barriers to extracting useful information when needed.

## 2.2. The LLM draft function in the DATE system

In response to this challenge, the DATE system introduced LLM draft functionality as an experimental feature during the first quarter of 2024 (January–March). This is part of an exploratory initiative aimed at understanding whether, and in what ways, LLM drafts can assist case handlers. By producing initial drafts of replies, this functionality seeks to reduce case handlers' workloads, allowing them to focus on more nuanced and critical aspects of communication to better support stakeholders.

The task of LLMs is to search through the DATE system's extensive database of historical cases and quickly generate draft replies to new inquiries using relevant information from similar past cases. These drafts are generated using the Retrieval-Augmented Generation (RAG) approach, which enhances the model's ability to create contextually relevant responses by integrating document retrieval with text generation. This approach ensures that the generated drafts are grounded in relevant historical records while being tailored to the specifics of the new inquiry.

The RAG method augments the input to the LLM with relevant information from knowledge sources (Lewis et al. 2020). It is known to enhance the accuracy and reliability, and thereby trustworthiness, of its output by enabling the LLM to reference credible knowledge sources outside of its training data when generating text. Research suggests that RAG approach provides improved control over LLM-generated content while reducing hallucinations, a common issue where LLMs produce incorrect, false, or fabricated information (G. Gao et al. 2024; Béchard and Ayala 2024; Shuster et al. 2021; Fan et al. 2024) .

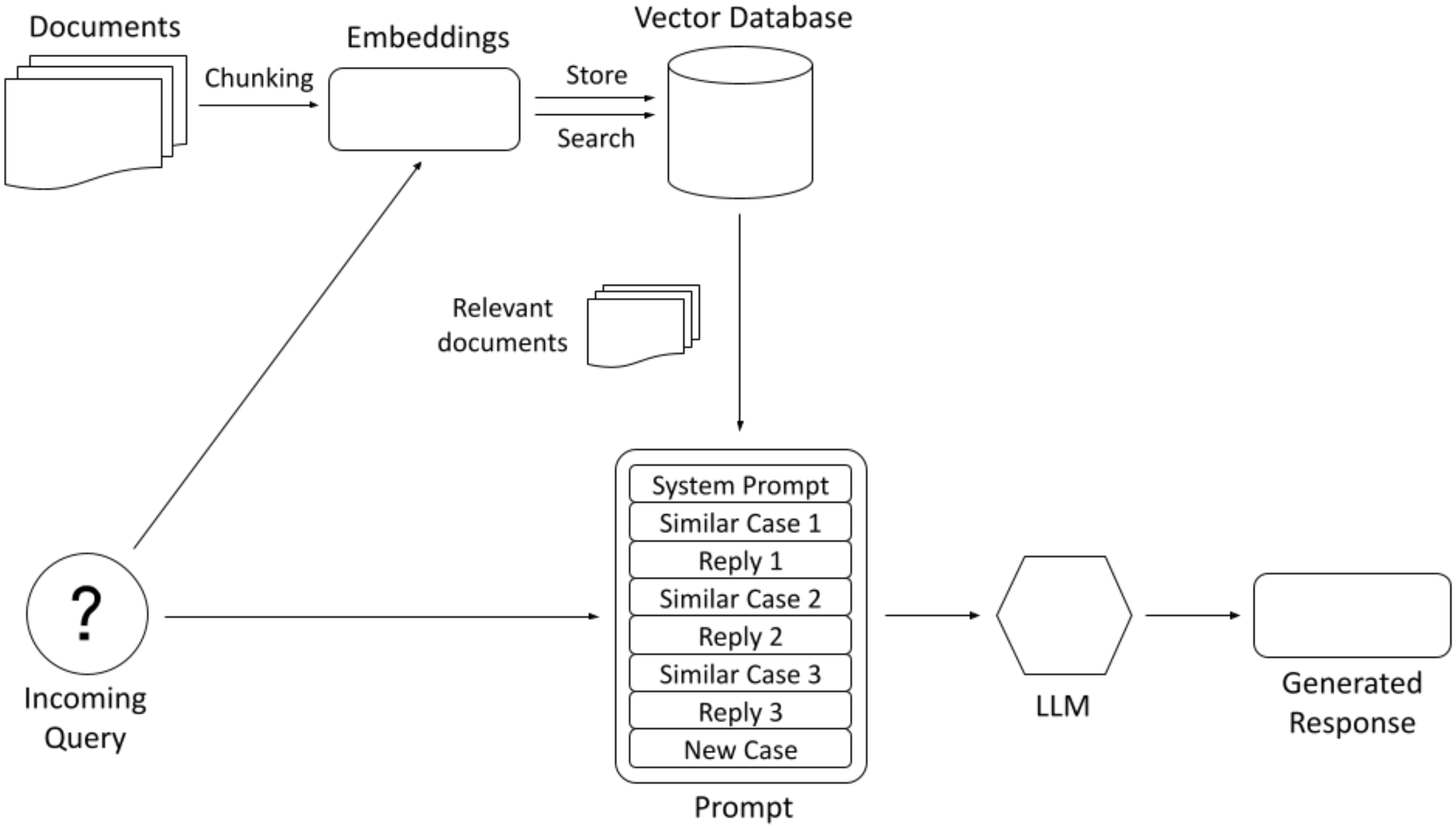

Figure 1. LLM draft generation process using Retrieval-Augmented Generation (RAG) in the DATE system. The system has a vector database of past cases. When a new query comes in, it retrieves the most similar cases from the vector database by using the vectorized query. The system uses these similar cases along with the new query to prompt the LLM for generating a contextually relevant response.

The process of generating draft responses to stakeholder inquiries using RAG in the DATE system is as follows (Figure 1): The DATE system maintains a vector database of 600,000 past inquiry cases. Existing inquiry cases are converted into text embedding vectors, numerical representations of the semantic aspects of the texts. When a new stakeholder inquiry is received, the system converts the inquiry text into an embedding vector and then searches the vector database to identify the most semantically similar cases using cosine similarity. The cases corresponding to these vectors are considered the most relevant to the new inquiry. The system then provides the LLM with a prompt that includes the three most similar inquiry-response pairs along with the new inquiry text. The LLM references these inquiry-response pairs as contextual examples to generate a response for the new inquiry. This allows LLM to generate a more contextually appropriate and accurate response for the new inquiry.  The LLM used for generating draft replies is OpenAI's ChatGPT-3.5 model.

## 2.3. LLM draft usage in practice

The LLM draft generation feature is currently in trial and optional for case handlers to use on a case-by-case basis. Case handlers are encouraged to explore the function critically. As part of our study, we disclosed to case handlers that the drafts were AI-generated, since the LLM was being co-developed with them and their feedback was essential to its improvement. When an LLM draft is generated, case handlers are also presented with references and links to similar cases that inform the draft, allowing direct access to these cases for further review. To receive the LLM drafts, case handlers must click the "Generate LLM Drafts" button in the user interface. It is important to note that this feature is only available for responding to a new inquiry; if multiple exchanges occur (e.g., an email chain), the feature is no longer accessible (Appendix A shows the distribution of the number of exchanges between the stakeholders and the case handlers per case). Another limitation is that attached documents (e.g., PDFs or images) are not included in the LLM's analysis. As a result, some crucial information may not be fully conveyed when generating drafts. In our setup, case handlers could, in principle, manually add relevant details from such documents into the inquiries if they wanted the LLM to take them into account, but this was not explicitly encouraged in the trial and was not a core part of the workflow we studied. Instead, the focus was on evaluating the usefulness of the draft generation as designed.

Case handlers can re-generate the LLM drafts as many times as they like, combining text from different versions before finalizing and sending the response to the stakeholders. Each time case handlers re-generate the LLM drafts, the newly generated text is displayed below the previous versions, allowing comparisons between versions.

# 3. Methodology

Our case study employed both subjective and computational measures to evaluate the usefulness of LLM drafts by case handlers.

**0. Preliminary study**

When the LLM draft functionality was first introduced, we conducted a preliminary qualitative study involving direct observations and interviews with case handlers. This initial phase aimed to gather early insights into how human experts responded to the new feature and its potential impact on their workflows. The findings from this qualitative study informed the design of the questionnaire used in the subsequent survey, ensuring it captured relevant themes and issues. Additionally, these findings provided a deeper contextual understanding of the case under study, improving the overall research process.

**1. A survey for case handlers (subjective measure)**

The goal was to capture the case handlers' perceptions of the LLM drafts (See Appendix B for the questionnaire). We disseminated the questionnaire between May and June 2024 to match the data collection timeline for our text similarity analysis.

**2. A text similarity analysis (computational measure)**

We measured the textual similarity between LLM drafts and sent replies. Our underlying assumption was that smaller textual differences would indicate that case handlers were able to make more effective use of the LLM drafts. To evaluate textual similarity, we used two methods as a triangulation approach: (1) semantic embedding similarity (SES), and (2) LLM-as-a-judge (LAAJ) rating. We also analyzed text similarity scores between SES and LAAJ to evaluate the reliability of results across LAAJ and SES through cross-verification and assess their complementarity. Data used for the analysis was collected for inquiry responses between May and June 2024 to match the survey data collection timeline.

## 3.1. Ethical considerations for human participation

Our case study was conducted in accordance with the guidelines provided by Sikt, the Norwegian Agency for Shared Services in Education and Research, a public administrative body under the Ministry of Education and Research in Norway. Based on Sikt's criteria, our research did not require formal ethical review for approval or notification, as it exclusively involved the collection and analysis of anonymous data and no personal data was processed (Sikt - Norwegian Agency for Shared Services in Education and Research). This means that at no point during the study – which included observations, interviews, a survey, and text similarity analysis – was it possible to identify individual participants. Thus, our research complied fully with Sikt's requirements and ensured the protection of participants' data (See Supplementary Material). In addition to not collecting personally identifiable information, all responses were recorded and stored in a manner that prevented the identification of individual participants.

All interview and observation participants gave informed consent prior to their participation and written informed consent was obtained. The process for interview and observation participants was thorough. Participants were first approached by email and received a detailed attachment outlining their rights as participants. At the start of every interview or observation session, participants were reminded that participation was entirely voluntary. Acceptance to participate was formally documented. Furthermore, participants were reminded at the conclusion of the session that they retained the right to withdraw any data gathered, at any time, by contacting the research team.

For survey participants, implied consent (a form of verbal consent) was obtained by their voluntary action of completing and submitting the questionnaire. This approach was chosen over written consent for the survey to ensure anonymity, maximize the convenience for a large sample, and maintain a high response rate, as the study presented no more than minimal risk to participants. Participation in the study was entirely voluntary, and all participants were informed that they could withdraw at any time without providing a reason. Participants were also informed that their data would be used exclusively for research purposes and handled in compliance with ethical and data protection standards. More detailed ethical considerations are also included in the next subchapters.

## 3.2. The survey study

### 3.2.1. The development of the questionnaire

To assess case handlers' experiences as users with the LLM drafts, a questionnaire was developed. The instrument was informed by in-depth discussions among the research team, DATE developers, and several case handlers to identify key aspects. In addition, several questionnaire items were adapted from established scales, including those by Brooke (1996), Davis (1989), and Venkatesh et al. (2003). This approach was to ensure that the questionnaire captured a comprehensive and relevant range of user perceptions and experiences while building upon existing measures. The questionnaire was tested on a few case handlers and iterated until all authors approved the final version.

The questionnaire consisted of 21 questions, in which 7 were categorized as demographic questions and 14 were measuring case handlers' experiences and opinions with the LLM drafts. The last demographic question measured how long a participant had been using the LLM drafts (i.e., a few days, a few weeks, a few months, or never). The survey was ended for those who responded with "never used the LLM drafts" after asking them to provide a reason (i.e., unavailable to them, chose not to use them, or other reasons followed by a free text comment field). Participants who had used the LLM drafts received 14 questions (see Appendix B). For these participants, the questionnaire was supposed to take 3-5 minutes to complete. The questionnaire was designed this way to minimize the time taken from case handlers' chargeable work hours.

### 3.2.2. Survey dissemination and participation

To ensure maximum participation, the questionnaire (Appendix B) was distributed electronically to a randomly selected sample of 198 case handlers within the DATE systems department based in USA, Norway, Greece, Singapore, Germany, and Poland. The selected sample population included case handlers with diverse roles, experience levels, and geographical locations. This ensured that the survey results reflected a broad range of perspectives and were representative of the organization's operational diversity. We adopted a sampling approach in alignment with organizational confidentiality policies to ensure that sensitive workforce data remained protected.

Prior to dissemination, we obtained formal approval from senior management to distribute the questionnaire. The link to the questionnaire was emailed directly to case handlers during May and June 2024 with a description and purpose of the questionnaire. We also sent three reminders during the data collection period. To encourage participation, the senior managers of DATE systems actively promoted the survey during department meetings, emphasizing its importance for improving case handling processes. No incentives were given to participants.

For the survey data collection, this case study complied with required ethical and legal standards relating to participant anonymity and confidentiality. Participation in the survey was anonymous and voluntary. The questionnaire was created and all responses were collected on Microsoft Teams Form with restricted access. Only the research team has access to the Teams channel and only the lead researcher of this case study had access to the questionnaire anonymous responses.

### 3.2.3. The analysis of the questionnaire responses

The independent variables included demographic questions and one question about trust (See Table 2 and Appendix B):

- The demographic questions: years of experience as a case handler, office location, age group, gender, current role, time using the LLM drafts, and native language(s)
    - The native language(s) were converted into non-native English speakers and native English speakers.
- The questionnaire item #7 "Stakeholders' trust in the organization": "I believe that the LLM drafts will: (1) increase, (2) decrease, or (3) have no impact on our stakeholders' trust in the organization".

The dependent variables were the questionnaire's items number 1-8 and 11 from Table 2 (all used a 4-point Likert-type scale). These questionnaire items of dependent variables were averaged for each participant as a total mean score rather than as individual questionnaire items. See Table 2 for selection of independent and dependent variables.

Table 2. The questionnaire items, response options, and what is being measured (See Appendix B for the full version of the questionnaire)

| Items | Response options | What is being measured |
|---|---|---|

| | | | |
|---|---|---|---|
| 1.<br>3.<br>4. | The LLM drafts are accurate.**<br>The language in the LLM drafts is of high quality.**<br>I find the summary of the LLM drafts useful.** | 4-point Likert-type scale from Strongly Disagree through to Strongly Agree | Perceived quality of the LLM drafts |
| 2.<br>5.<br>9.<br>11. | I would recommend the LLM drafts to other case handlers.**<br>My colleagues and I discuss the use of the LLM drafts frequently.**<br>I intent to continue using the LLM drafts in the next 12 months.**<br>I use the LLM drafts frequently.** | 4-point Likert-type scale from Strongly Disagree through to Strongly Agree | Intent to use and recommend the LLM drafts |
| 12.<br>10.<br>8. | Using the LLM drafts increases my productivity (1).**<br>Since I started using the LLM drafts, I consult with my colleagues: [less often/more often/about the same] (2).<br>Using the LLM drafts improves my responses to stakeholders in terms of [clarity, accuracy, efficiency, consistency, conciseness, stakeholder satisfaction, professionalism, and/or other [free text]] (3). | (1) 4-point Likert-type scale from Strongly Disagree through to Strongly Agree<br>(2) 3-point Likert-type scale<br>(3) Multiple responses possible (seven factors and one free text) | Perceived impact of the LLM drafts on own work performance |
| 6.<br>13. | I tend to modify the LLM drafts: (1).**<br>When in doubt, I rely on [the LLM drafts, my knowledge, and/or my colleagues] (2). | (1) 4-point Likert-type scale: Extensively, moderately, slightly, none at all***<br>(2) Multiple responses possible | Dependency and modification on the LLM drafts |
| 7. | I believe that the LLM drafts will [increase/decrease/have no impact] on our stakeholders' trust in the organization.* | 3-point Likert-type scale | Impact of the LLM drafts on the stakeholders' trust |
| 14 | General comments or suggestions about the LLM drafts. | Free text comments | Open-ended feedback |

*This questionnaire item is treated as an independent variable for statistical analyses
** This questionnaire item is treated as a dependent variable for statistical analyses
***Extensively is scored as Strongly Disagree, Moderately as Disagree, Slightly as Agree, and None at all as Strongly Agree

We compared the total scores for the independent variables (i.e., demographic information and the impact on stakeholders' trust in the organization) to the dependent variables (i.e., the total mean scores of the nine questionnaire items) using Kruskal-Wallis H test. This determined whether there were statistically significant differences in the total scores between groups for: the current roles, years of experience being a case handler, office locations, non-native vs. native English speakers, age groups, gender, length of using the LLM drafts, and the impact on stakeholders' trust in the organization. We conducted post-hoc pairwise comparisons using Dunn's test with a Bonferroni correction applied to control for multiple comparisons. This adjusted the significance level to $p < 0.017$.

We tested differences in the composition of the demographic information between users (i.e., case handlers who had used the LLM drafts) and non-users (i.e., case handlers who never used the LLM drafts) using Pearson's chi-square tests separately within each demographic (i.e., the current roles, years of experience being a case handler, office locations, non-native vs. native English speakers, age groups, gender, length of using the LLM drafts, and the impact on stakeholders' trust in the organization) as factors. If interactions of group and category were found (i.e., if the demographic composition of users and non-users differed), we conducted follow-up pairwise Pearson's chi-square tests between sub-groups. Odds ratios (where an odds ratio of 1 indicates that the demographic composition of users and non-users did not differ) calculated the effect size. We used a significance level of $p < 0.05$, the Bonferroni correction for each family of comparisons, where necessary. In addition, we computed Pearson correlation coefficients between the nine questionnaire items.

In addition, we conducted a thematic analysis following an inductive approach of the survey response free text for the questionnaire item #14 (i.e., General comments or suggestions about the LLM drafts: [free text]). We identified and extracted significant words and phrases from each transcript and interpreted their meanings in context. Codes expressing similar meanings were grouped into themes.

## 3.3. Text similarity analysis

The initial dataset for text similarity analysis included all 13,794 stakeholder inquiries received through the DATE system between May and June 2024 (Figure 2).

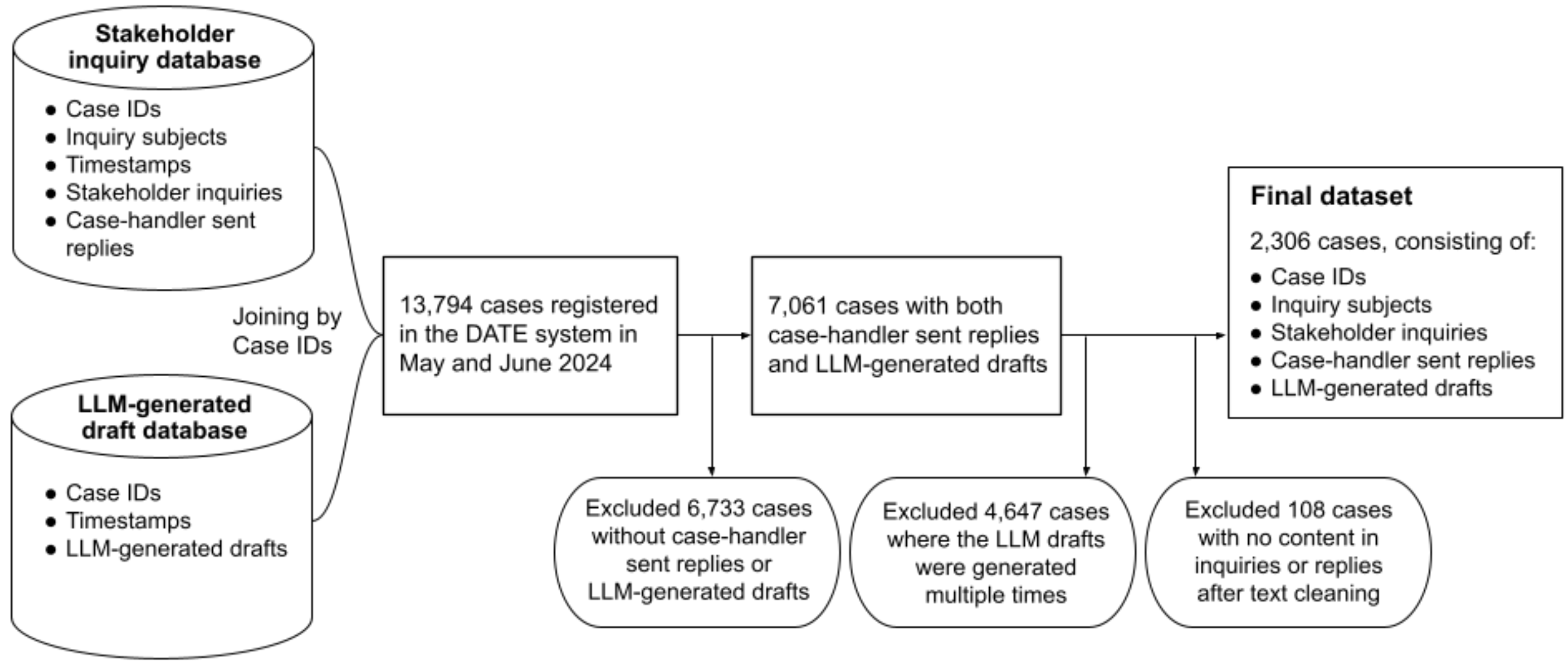

Figure 2. Dataset construction workflow for the text similarity analysis.

The DATE system stores stakeholder inquiries and LLM-generated drafts in separate databases. We merged these using case IDs to create a dataset linking drafts with the corresponding replies sent by case handlers. This process yielded 7,061 cases after excluding 6,733 cases where either the reply or the draft was missing.

To ensure comparability, we further restricted the dataset to 2,414 cases in which only one LLM draft was generated, excluding 4,647 cases with multiple drafts (see Appendix C for distribution). Multiple draft generations risked case handlers combining text from different versions, making it difficult to identify the specific draft associated with a reply. Although this restriction may not capture all usage patterns, it enables clear draft–reply matching and improves the accuracy and consistency of the analysis.

To protect personal information and focus on the inquiry content, personally identifiable details (e.g., sender and recipient names, salutations, affiliations, and contact information) were removed using automated algorithms. To standardize text formats and prevent non-content elements (e.g., HTML tags, line breaks, and consecutive spaces) from affecting text comparison, these elements were also removed from the dataset using automated algorithms. After cleaning, 108 cases where inquiries or replies were left without content were excluded, resulting in a final dataset containing 2,306 cases.

### 3.3.1. Text Similarity Measures

To computationally compare LLM drafts and replies sent by case handlers, we employed two measures of text similarity: semantic embedding similarity (SES) and LLM-as-a-judge (LAAJ). We then compared the similarity scores produced by these two measures to evaluate their consistency and reinforce the reliability of our findings.

#### *Semantic embedding similarity (SES)*

Semantic embedding represents text as a vector in a high-dimensional space, where each dimension corresponds to a latent semantic feature learned by the model during its training on vast text corpora (Selva Birunda and Kanniga Devi 2021; Bojanowski et al. 2017; T. Gao, Yao, and Chen 2021; Mikolov 2013; Pennington, Socher, and Manning 2014). This vector captures semantic nuances, allowing for the quantification of semantic similarity between two texts by measuring the closeness between their respective vectors. Cosine similarity is a common metric used for this purpose, quantifying how close two text embeddings are by calculating the cosine of the angle between their vectors (Reimers 2019; J. Wang and Dong 2020). For text embeddings, cosine similarity scores typically range from 0 to 1, where 1 indicates perfect similarity and 0 indicates no similarity.

For each of the 2,306 cases in the final dataset, we converted both the LLM-generated text (the LLM draft) and the case handler-generated text (the sent reply) into semantic embedding vectors and calculated the cosine similarity between them. This process yielded a semantic embedding similarity score for each case, as well as the overall similarity score distribution across the dataset. For converting text into semantic embedding vectors, we used two OpenAI text embedding models to ensure the reliability of the results: text-embedding-ada-002 and text-embedding-3-large[1].

#### *LLM-as-a-judge (LAAJ)*

LAAJ leverages the advanced language comprehension capabilities of LLMs to evaluate the quality of texts based on specific instructions (prompts). This method is grounded in LLMs' extensive language knowledge and contextual understanding abilities, enabling detailed and specialized evaluations across various dimensions of language. Many recent studies have actively discussed the use of LAAJ for computational text evaluation (Bavaresco et al. 2024; Chiang and Lee 2023; Laskar, Chen, and Tn 2023; Yang Liu et al. 2023; Sperber 2004; Zheng et al. 2023).

---

[1] The text-embedding-ada-002 model, released in December 2022, has been widely used for various natural language processing tasks. It generates 1536-dimensional embeddings and has demonstrated robust performance across a range of applications. The more recent text-embedding-3-large model, introduced in January 2024, features up to 3072 dimensions, enhancing performance compared to its predecessors (https://openai.com/index/new-embedding-models-and-api-updates/)

We adopted three evaluation criteria for the LLM's assessment: comparability of language, similarity of interpretability, and overall similarity. Language comparability refers to the formal similarity of words, phrases, and sentences between texts, while similarity of interpretability assesses the degree to which two versions evoke the same response even when their wording differs (Sperber 2004). By also having the LLM evaluate the overall similarity between the LLM draft and the sent reply, we established a comprehensive framework for detailed analysis. The LLM's evaluations were conducted using a 7-point Likert scale, where 1 represented "not at all comparable/similar" and 7 represented "highly comparable/similar." Additionally, to enhance transparency and enable researchers to verify the evaluations, we instructed the LLM to provide explanations justifying its assessments. The complete prompt provided to the LLM, including these instructions, can be found in Appendix D. We used OpenAI's GPT-4o (version 2024-05-13) as the LLM for LAAJ. We utilized Microsoft Azure OpenAI services to implement our evaluation framework and designed the system to generate and store structured responses in JSON format for easier analysis.

For each of the 2,306 cases in the final dataset, we provided the LLM with the prompt along with the LLM draft and the sent reply to evaluate textual similarity. As a single metric for comparison with other similarity measures like SES, we created a composite score by taking the mathematical average of the three dimension-specific scores obtained (i.e., overall similarity, language comparability, and similarity of interpretation), representing combined evaluations across all three dimensions. Furthermore, to mitigate potential variability from the LLM's probabilistic nature, we conducted three independent evaluation runs and averaged these scores to obtain more stable and representative final similarity scores.

## *Comparison of SES and LAAJ Scores*

As a methodological triangulation approach for text similarity assessment, we conducted comparative analysis of the similarity scores obtained through both SES and LAAJ methods. We calculated the correlations between SES and LAAJ similarity scores across the entire dataset using Pearson and Spearman correlation coefficients. We also examined the statistical characteristics and density distribution plots of both similarity score distributions to thoroughly compare and analyze the distribution patterns of the two similarity scores.

# 4. Results

## 4.1. The preliminary findings

To complement our main survey and text similarity analysis, we conducted a set of preliminary, exploratory activities aimed at providing contextual insights into how the LLM system was used in practice. These activities included a total of 42 hours of direct, non-participant observations, four semi-structured interviews with case handlers, and two semi-structured interviews with developers.

The observations focused on case handlers' day-to-day work, particularly how they interacted with LLM-generated drafts, integrated them into replies, and navigated supporting systems. Field notes captured examples of challenges (e.g., alignment between draft suggestions and workflow) and opportunities (e.g., efficiency in searching for previous cases).

The interviews with case handlers provided reflections on the practical usefulness of the drafts, factors influencing adoption, and perceptions of accuracy and relevance. The developer interviews offered complementary perspectives on the design choices, system constraints, and expectations regarding draft uptake.

All notes and transcripts were coded thematically, with an emphasis on identifying recurrent patterns rather than statistical generalization. These findings are presented as illustrative and exploratory: their purpose is to shed light on possible mechanisms behind the patterns observed in the main analyses, rather than to serve as stand-alone results.

**Levels of complexity of inquiries**

Stakeholder inquiries present an almost limitless range of variations, scenarios and combinations. These include factors such as vessel type, construction site, operational timeline, sailing conditions, geographical location, local regulations, crew history, vessel maintenance records, engineering culture, maintenance decisions, shipping tax regulations, onboard equipment types, and the diversity of equipment suppliers, among others. This made users question whether an LLM could have been accurately trained on such complex data. We also found that the level of complexity of stakeholder inquiries significantly influenced how case handlers perceived the usefulness of LLM drafts. A case handler's quote illustrates perceived challenges for the LLM to address complex inquiries:

> *"In the maritime industry, every ship is different, even sister ships have differences. The problems they encounter are variable. This is because different people manage the ship, different management styles, different operational areas, cargoes, and flags. ... So probably [the LLM] would take the answer from an already answered question, which is of similar age, vessel, similar type of vessel, similar flag, but not necessarily the same management." – A case handler*

Conversely, simpler inquiries (e.g., those involving clear procedures and straightforward instructions with readily available explicit knowledge) tended to inspire greater trust in the LLM drafts. In such cases, the required information was often readily available in past cases or documented sources like regulations and procedures, making it easier for the LLM to generate more accurate LLM drafts. Case handlers also suggested that these low-complexity cases would be easier for them to verify.

**Knowledge sharing**

Case handlers dealt with the level of complexity presented by the stakeholder inquiries by discussing with colleagues, calling surveyors on board for field-based information, and using their experiences and domain knowledge. Our observations also suggest a high level of collaboration and teamwork between and among case handlers and other colleagues, especially when responding to complex inquiries. Often, handlers made decisions collaboratively, rather than alone, to raise the confidence and quality of the replies to the stakeholder inquiries. Case handlers also reported that collaboration and teamwork might also lead to the creation of new knowledge or new ways of dealing with cases. Lastly, case handlers felt that the usual way they responded to inquiries, especially complex ones, could not be incorporated into the LLM since this involved tacit knowledge.

**Brainstorming**

Although the LLM drafts were created to help case handlers respond to the inquiries, we observed that the LLM drafts were used in other ways as well. Case handlers leveraged LLM drafts to generate ideas for addressing new inquiries. They reflected on aspects, such as necessary information, relevant regulations, and applicable domain expertise, to help shape their replies. In addition, case handlers tested a filtering function within the user interface (UI) that enabled them to collaborate with the LLM in retrieving cases used to generate drafts, based on criteria such as vessel type, flag, or age. Case handlers explored how different filters changed the content of the LLM drafts to generate different ideas and evaluate the specificity of the content of the LLM drafts.

**Skeptical but curious**

Initially skeptical, case handlers carefully evaluated the LLM drafts to determine alignment with their knowledge. When discrepancies arose, they often dismissed the system as immature. However, they revisited it after updates or improvements were announced, demonstrating a cautious openness to innovation.

**Humanizing**

In the cases when case handlers did not understand the LLM drafts, they tended to humanize the technology by making comments such as calling it a toddler, that "it" did not understand the context of the inquiry, and "it" made up content.

**Trust**

We observed that the trust of case handlers in technology increased when they recognized phrases from prior cases. In contrast, LLM drafts perceived as drawing primarily from generic internet sources were met with greater skepticism.

## 4.2. The survey results

A total of 74 of 198 invited, selected case handlers responded to the questionnaire (response rate = 37.37%), in which 23 selected "never used the LLM drafts," leaving 51 participants (response rate = 25.76%) to respond to the 14 questionnaire items measuring their experiences and opinions of the LLM drafts (Figure 3). In this case study, we designated participants who reported to have used the LLM drafts as "users" and those who reported to never use the LLM drafts as "non-users". Non-users commonly reported their reasons for never using the LLM drafts were that it was not fit-for-purpose and unavailable (Table 3). Unavailability to non-users was probably because the non-users moved to another unit or team without access to the LLM draft function on trial. It might also be possible that non-users were not aware of the new functionality.

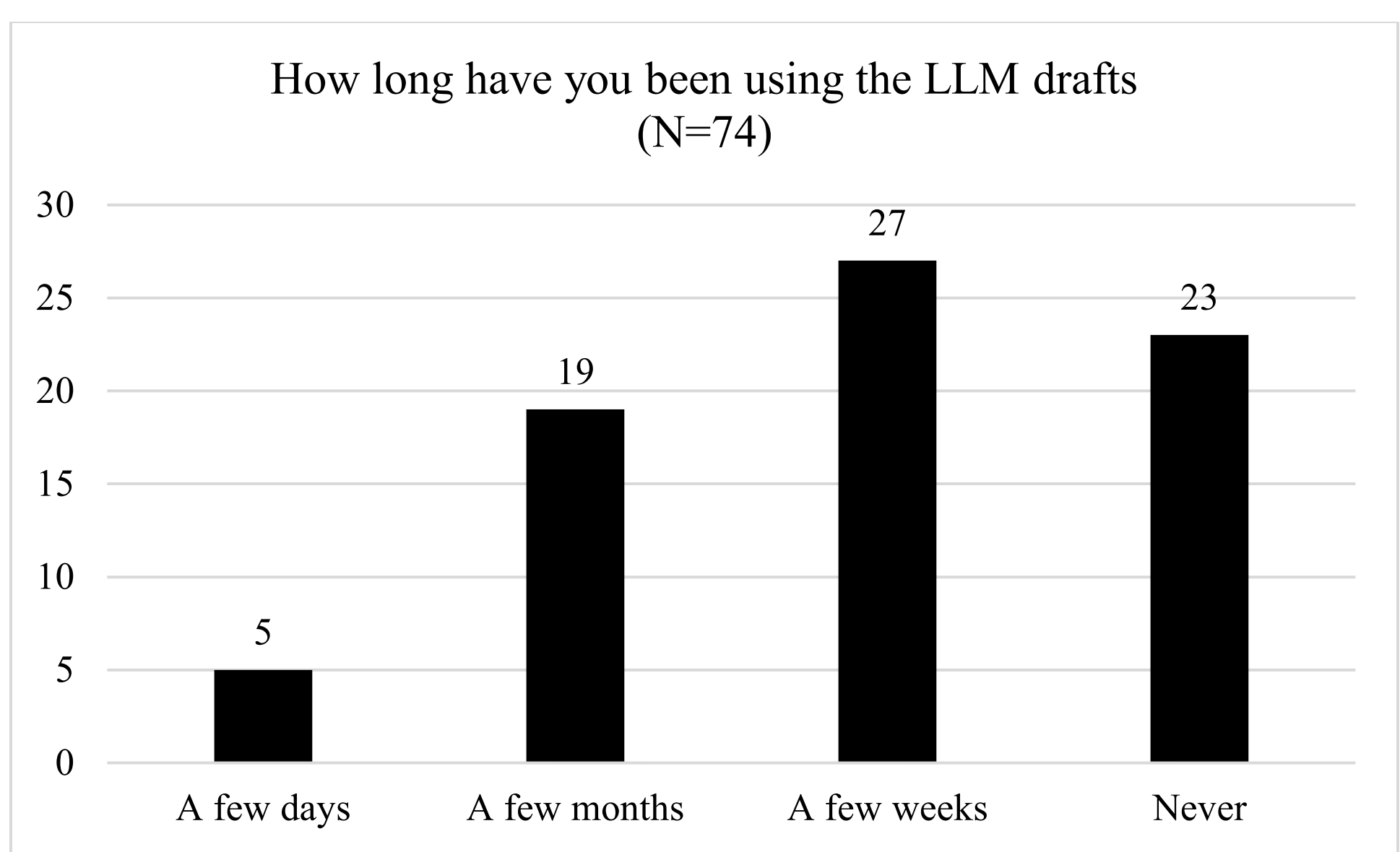

Figure 3. Participant distribution of users (had used the LLM drafts) and non-users (never used the LLM drafts)

Table 3 Reasons cited by non-users for never utilizing the LLM drafts

| Reasons | N | Summarized free text responses |
|---|---|---|
| Unavailable to the non-users | 6 | |
| Available, but non-users chose not to use the LLM drafts* | 5 | • The LLM drafts were perceived as too generic, often incorrect or incomplete, and not helpful for specialized, specific, and urgent cases<br>• Non-users perceived that modifying third-party LLM drafts required more effort than creating new texts from scratch. |
| Other reasons | 12 | • Non-users were still in the process of familiarizing themselves with the LLM drafts, had not yet taken the time to start this process, or forgotten about the LLM drafts.<br>• The LLM drafts were perceived as too generic, irrelevant or not applicable for most cases, especially for specialized, specific, and urgent cases.<br>• The perception that writing replies to stakeholder inquiries manually was faster than using the LLM drafts.<br>• Stakeholders sent their inquiries via other channels such as emails. |

*Non-users who selected this response received a follow-up question to provide their reasons in free texts.

Table 4 shows the demographic information of all the participants, users, and non-users. The most common role of the case handlers across the three groups was engineering. All participants and users most commonly had 5-10 years experience as case handlers, whereas non-users most commonly had longer than 15 years' experience. Norway and Germany were the most common office locations for the participants across the three groups. The most common age group for participants across the three groups was older than 50 years, followed by 41-50 years old. Most of the participants identified themselves as male across the three groups. German and Norwegian were the most common native languages across the three groups. Similarly, most of the participants across the three groups were categorized as non-native English speakers. A Pearson's chi-square analysis (N = 51) revealed that there were no statistically significant differences in the composition of the demographic information between users and non-users (Table 5).

Table 4. Demographic information of all participants, users, and non-users

| Category | | All participants (N=74) | | Users (N=51) | | Non-users (N=23) | |
|---|---|---|---|---|---|---|---|
| | | **N** | **%** | **N** | **%** | **N** | **%** |
| Current role | Engineer | 39 | 52.70 % | 26 | 50.98 % | 13 | 56.52% |
| | Support | 13 | 17.57 % | 10 | 19.61 % | 3 | 13.04% |
| | Surveyor | 11 | 14.86 % | 9 | 17.65 % | 2 | 8.70% |
| | Specialist | 6 | 8.11 % | 2 | 3.92 % | 4 | 17.39% |
| | Management | 4 | 5.41 % | 4 | 7.84 % | 0 | 0 |
| | Consultant | 1 | 1.35 % | 0 | 0 | 1 | 4.35% |
| | **TOTAL** | **74** | **100 %** | **51** | **100%** | **23** | **100%** |
| Years of experience as a case handler | <1 year | 4 | 5.41 % | 4 | 7.84 % | 0 | |
| | 1-5 years | 19 | 25.68 % | 14 | 27.45 % | 5 | 21,74 % |
| | 5-10 years | 23 | 31.08 % | 18 | 35.29 % | 5 | 21,74 % |
| | 10-15 years | 13 | 17.57 % | 8 | 15.69 % | 5 | 21,74 % |
| | >15 years | 15 | 20.27 % | 7 | 13.73 % | 8 | 34,78 % |
| | **TOTAL** | **74** | **100 %** | **51** | **100%** | **23** | **100%** |
| Office location | Norway | 31 | 41.89 % | 17 | 33.33 % | 14 | 60.87% |
| | Germany | 23 | 31.08 % | 15 | 29.41 % | 8 | 34.78% |
| | USA | 8 | 10.81 % | 8 | 15.69 % | 0 | 0 |
| | Singapore | 6 | 8.11 % | 6 | 11.76 % | 0 | 0 |
| | Greece | 3 | 4.05 % | 3 | 5.88 % | 0 | 0 |
| | Poland | 3 | 4.05 % | 2 | 3.92 % | 1 | 4.35% |
| | **TOTAL** | **74** | **100 %** | **51** | **100%** | **23** | **100%** |
| Age group | 20-30 years old | 4 | 5.41 % | 3 | 5.88 % | 1 | 4.35 % |
| | 31-40 years old | 7 | 9.46 % | 7 | 13.73 % | 0 | 0 % |
| | 41-50 years old | 25 | 33.78 % | 17 | 33.33 % | 8 | 34.78 % |
| | >50 years old | 38 | 51.35 % | 24 | 47.06 % | 14 | 60.87 % |
| | **TOTAL** | **74** | **100 %** | **51** | **100%** | **23** | **100%** |
| Gender | Female | 11 | 14.86 % | 5 | 9.80% | 6 | 26.09 % |
| | Male | 60 | 81.08 % | 43 | 84.31% | 17 | 73.91 % |
| | Prefer not to say | 3 | 4.05 % | 3 | 5.88% | 0 | 0 % |
| | **TOTAL** | **74** | **100 %** | **51** | **100%** | **23** | **100%** |
| Native language(s) | German | 19 | 25.68 % | 13 | 25.49 % | 6 | 26.09 % |
| | Norwegian | 18 | 24.32 % | 10 | 19.61 % | 8 | 34.78 % |
| | English | 8 | 10.81 % | 6 | 11.76 % | 2 | 8.70 % |
| | Chinese/Mandarin | 6 | 8.11 % | 6 | 11.76 % | 0 | 0 % |
| | Greek | 4 | 5.41 % | 3 | 5.88 % | 1 | 4.35 % |
| | Polish | 3 | 4.05 % | 2 | 3.92 % | 1 | 4.35 % |
| | No information | 2 | 2.70 % | 2 | 3.92 % | 0 | 0 % |
| | Bengali | 2 | 2.70 % | 1 | 1.96 % | 1 | 4.35 % |
| | Spanish | 2 | 2.70 % | 2 | 3.92 % | 0 | 0 % |
| | French | 2 | 2.70 % | 1 | 1.96 % | 1 | 4.35 % |
| | Turkish | 2 | 2.70 % | 2 | 3.92 % | 0 | 0 % |
| | Cantonese | 1 | 1.35 % | 1 | 1.96 % | 0 | 0 % |
| | Dutch | 1 | 1.35 % | 1 | 1.96 % | 0 | 0 % |
| | Greek & German | 1 | 1.35 % | 0 | 0 | 1 | 4.35 % |
| | Hindi | 1 | 1.35 % | 1 | 1.96 % | 0 | 0 % |
| | Russian | 1 | 1.35 % | 0 | 0 | 1 | 4.35 % |
| | Slovenian | 1 | 1.35 % | 0 | 0 | 1 | 4.35 % |
| | **TOTAL** | **74** | **100 %** | **51** | **100%** | **23** | **100%** |
| Native English speaker | Yes | 8 | 10.81 % | 6 | 11.76% | 2 | 8.70 % |
| | No | 62 | 83.78% | 43 | 84.31% | 21 | 91.30 % |
| | No information | 2 | 2.70% | 2 | 3.92% | 0 | 0 |
| | **TOTAL** | **74** | **100 %** | **51** | **100%** | **23** | **100%** |

Table 5. Results of Pearson's chi-squared tests for the composition of demographic information between users and non-users

| **Variable** | **Chi-Squared ($\chi^2$)** | **N** | **Degrees of Freedom (df)** | **p-value** | **Statistically significance** |
|---|---|---|---|---|---|
| Current Role | 8.904 | 74 | 5 | 0.113 | No |
| Years of Experience as a Case Handler | 6.740 | 74 | 4 | 0.150 | No |
| Office Location | 10.690 | 74 | 5 | 0.058 | No |
| Age Group | 3.825 | 74 | 3 | 0.281 | No |
| Gender | 4.392 | 74 | 2 | 0.111 | No |
| Native English Speakers or not | 1.130 | 74 | 2 | 0.569 | No |

Over 60% of the users (N=34) did not provide any responses when asked how the LLM drafts improved their responses to stakeholder inquiries. Of the 17 users (33.33%) who did respond (Figure 4), the majority indicated that the LLM drafts enhanced their efficiency. Almost all users (92.16%) stated that using the LLM drafts did not affect their frequency of consulting with their colleagues (Figure 5) Users relied mostly on their knowledge and colleagues rather than on the LLM drafts when in doubt (Figure 6). The nine users who selected 'Other,' either alone or in combination with other options, provided free-text responses indicating that, when in doubt, they relied on previous replies to similar cases and on rules and regulations. One user mentioned that they did not rely on the LLM drafts until they believed the technology had become more mature. More than half of the users (50.98%) believed that the LLM drafts would have no impact on the stakeholders' trust in the organization (Figure 7).

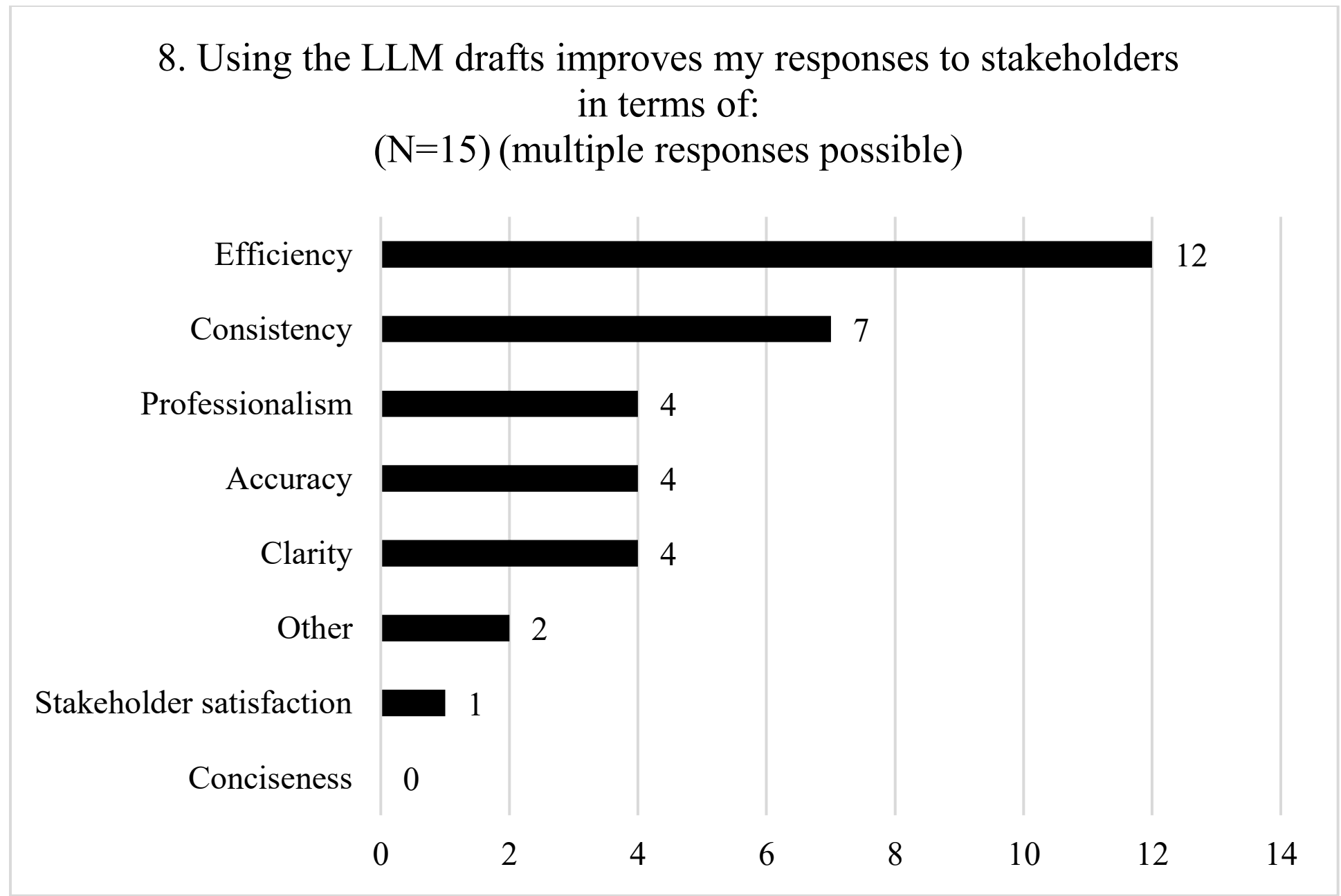

Figure 4. Users' feedback on how the LLM drafts improved their responses to inquiries included the following themes: 'Limited improvement' and 'Incorrect content' from two 'Other' free-text responses

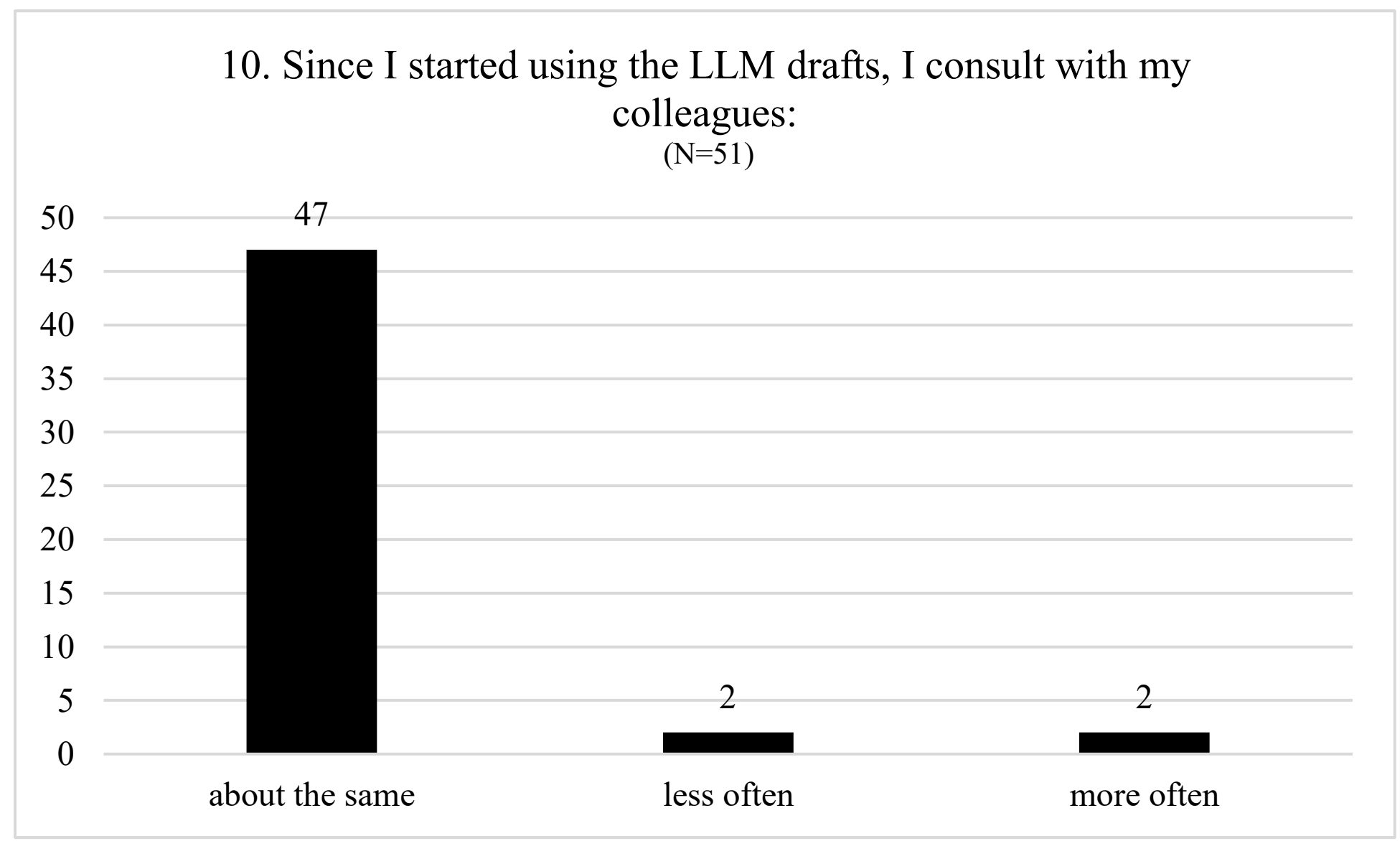

Figure 5. Users' responses to whether using the LLM drafts changed the frequency of them consulting with colleagues

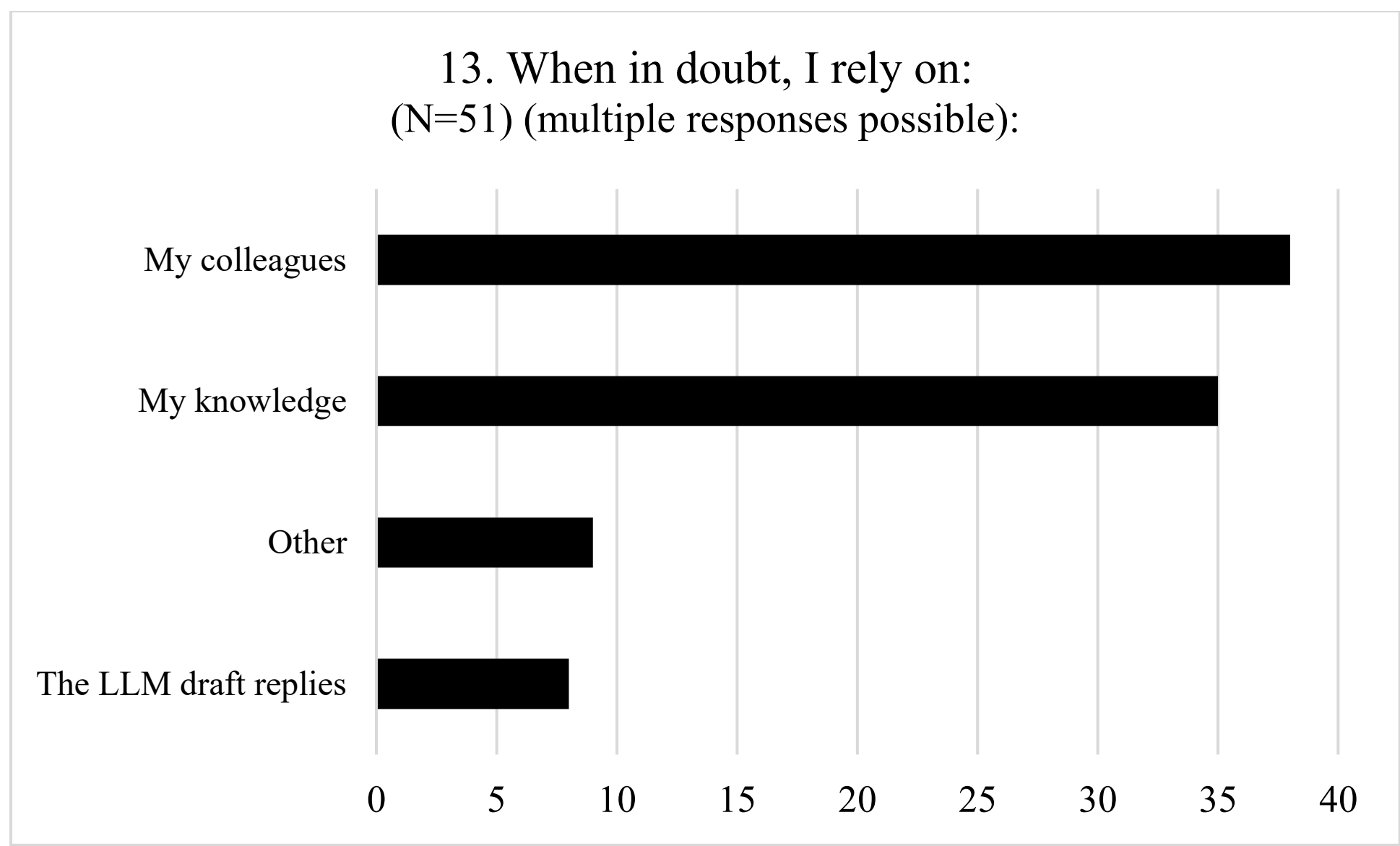

Figure 6. Users' responses to reliance sources when in doubt

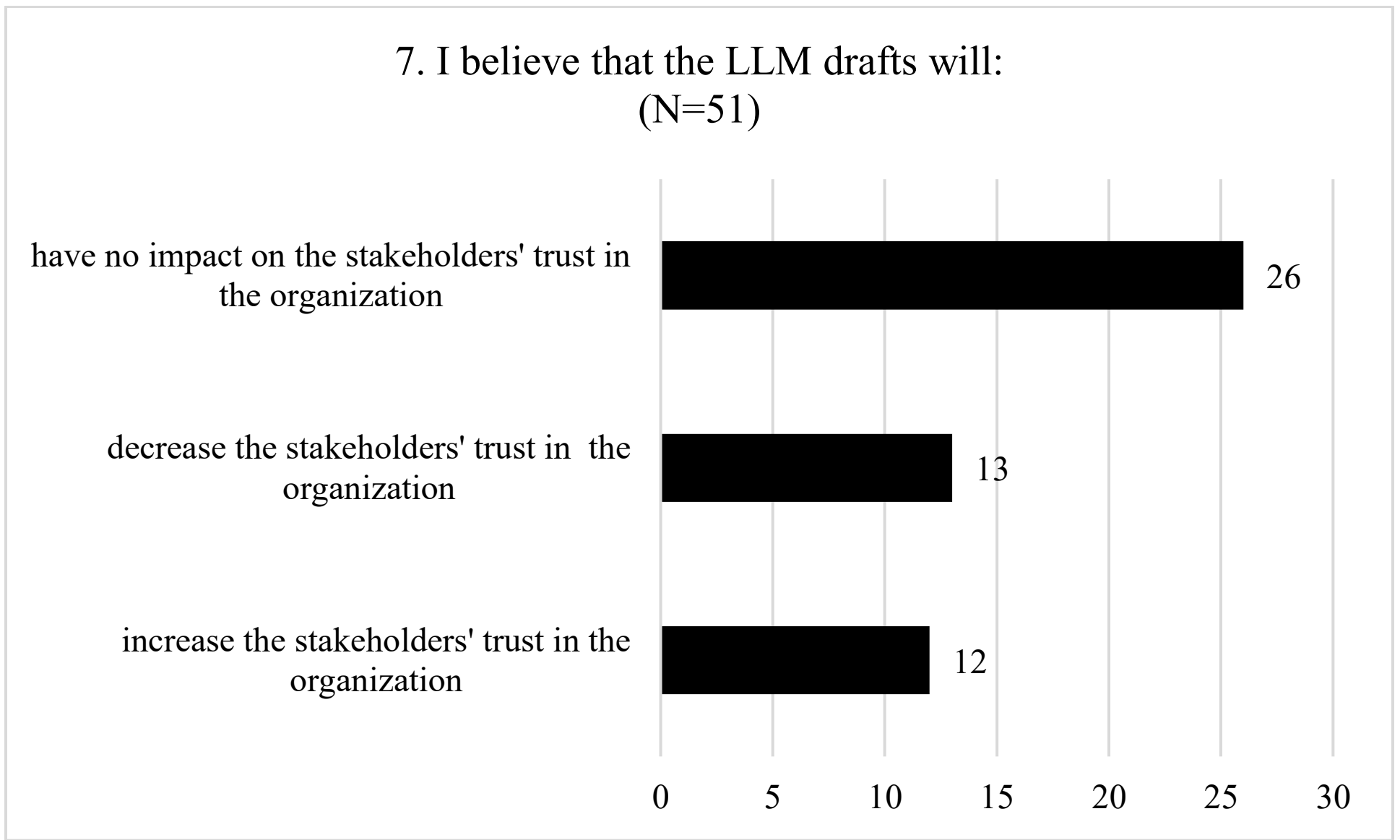

Figure 7. Users' responses to the impact of the LLM drafts on the stakeholders' trust in the organization

In general, users responded more negatively to items 1, 6, 8, and 11 (Table 6). This shows that they perceived that the LLM-generated draft replies' quality still needed significant improvement. Users responded more positively to the items 2, 3, 5, and 9, indicating that users intended to continue using the LLM drafts and recommend others use them as well. Users also perceived the language as of high quality. This shows that although the quality of the LLM drafts was perceived as an area for improvement, users were willing to continue using the LLM drafts. A split was observed whether the summary of the LLM drafts was perceived as useful. See Appendix E for correlation coefficient between the nine questionnaire items.

Table 6. Distribution of user responses to the nine questionnaire items measuring user perceptions of the LLM drafts

| **The questionnaire items measuring users' perceptions of the LLM drafts** | **Strongly Disagree** | | **Disagree** | | **Agree** | | **Strongly Agree** | | **Total** | |
|---|---|---|---|---|---|---|---|---|---|---|
| | **N** | **%** | **N** | **%** | **N** | **%** | **N** | **%** | **N** | **%** |
| 1. The LLM drafts are accurate. | 7 | 13.73 % | 37 | 72.55 % | 7 | 13.73 % | 0 | 0 % | 51 | 100 % |
| 2. I would recommend the LLM drafts to other case handlers. | 3 | 5.88 % | 19 | 37.25 % | 25 | 49.02 % | 4 | 7.84 % | 51 | 100 % |
| 3. The language in the LLM drafts is of high quality. | 1 | 1.96 % | 7 | 13.73 % | 35 | 68.63 % | 8 | 15.69 % | 51 | 100 % |
| 4. I find the summary of the LLM drafts useful. | 4 | 7.84 % | 21 | 41.18 % | 22 | 43.14 % | 4 | 7.84 % | 51 | 100 % |
| 5. My colleagues and I discuss the use of the LLM drafts frequently | 3 | 5.88 % | 17 | 33.33 % | 30 | 58.82 % | 1 | 1.96 % | 51 | 100 % |
| 6. I tend to modify the LLM drafts:* | 36 | 70.59 % (extensively) | 9 | 17.65 % (Moderately) | 3 | 5.88 % (Slightly) | 3 | 5.88 % (None at all) | 51 | 100 % |
| 8. Using the LLM drafts increases my productivity. | 3 | 5.88 % | 27 | 52.94 % | 19 | 37.25 % | 2 | 3.92 % | 51 | 100 % |
| 9. I intent to continue using the LLM drafts in the next 12 months. | 4 | 7.84 % | 7 | 13.73 % | 32 | 62.75 % | 8 | 15.69 % | 51 | 100 % |
| 11. I use the LLM drafts frequently. | 7 | 13.73 % | 27 | 52.94 % | 15 | 29.41 % | 2 | 3.92 % | 51 | 100 % |

*Extensively was scored as Strongly Disagree, Moderately as Disagree, Slightly as Agree, and None at all as Strongly Agree

The results from statistical analyses show that the only significant difference was the impact of the LLM drafts would have on the stakeholders' trust in the organization (Table 7). We conducted a Dunn's post-hoc test with Bonferroni correction to follow up on the significant Kruskal-Wallis test results for the variable "stakeholders' trust in the organization" on "responses." The results indicated significant differences between the "decrease" and "increase" groups, $Z = -4.29$, $p_{adj} < 0.001$, and between the "decrease" and "no impact" groups, $Z = -4.07$, $p_{adj} < 0.001$. There was no significant difference between the "increase" and "no impact" groups, $Z = 0.96$, $p_{adj} = 1.00$. The "decrease" group's mean score was 1.69 of 4.00, the "increase" group's was 2.00, and the "no impact" group's was 2.15. Based on these mean scores, the "decrease" group seemed to have the most negative perception of the overall score of the nine dependent variables.

Table 7. Results of Kruskal-Wallis statistical tests for differences in responses across users' demographic and background variables

| Variable | $\chi^2$ (df) | N | p-value | Significance |
|---|---|---|---|---|
| Years of experience as a case handler | 7.54 (4) | 51 | 0.11 | No |
| Office location | 1.69 (5) | 51 | 0.89 | No |
| Age groups | 4.49 (3) | 51 | 0.21 | No |
| Gender | 3.31 (2) | 51 | 0.19 | No |
| Current role | 5.88 (4) | 51 | 0.21 | No |
| Native English speaker or not | 1.84 (2) | 51 | 0.40 | No |
| Length of time using the LLM drafts | 5.30 (2) | 51 | 0.07 | No |
| Stakeholders' trust in the organization | 22.41 (2) | 51 | < 0.001 | Yes |

We conducted correlation analysis on the nine questionnaire items, resulting in 36 pairwise correlations (Appendix E). Of these, 20 correlations were statistically significant ($p < 0.05$), indicating meaningful relationships between the corresponding variables. All the significant correlations showed positive relationships, suggesting that as one variable increased, the other also tended to increase. In contrast, 16 correlations were not statistically significant ($p > 0.05$), indicating no strong evidence of a relationship between those pairs of items. This suggests that responses to these items are independent of each other.

The general comments from the questionnaire item #14 from the users about the LLM drafts were summarized as follows:

**1. Accuracy and tailoring of the LLM drafts**

Users often perceived the LLM drafts to lack technical accuracy and relevance, with some comments noting that replies were either too general, missing the point of the stakeholders' questions, or repeated the same information. Although users perceived the language of the LLM drafts as excellent, they suggested that the content lacked substantive content or even provided incorrect information. The need for the LLM to handle cases individually, rather than applying a one-size-fits-all approach, was a recurring theme.

Consequently, users expressed a need for the LLM drafts to be more precise and better tailored to the specific details of each case. They also wanted to allow for customization based on specific stakeholder needs and the complexity of the inquiries. This might be achieved by allowing the LLM access the documents attached to stakeholder inquiries and by considering regional specifics such as differences in handling, fees, and regulations in different countries.

Importantly, several comments emphasized the need for the LLM to integrate data from other sources and databases to provide more informed and contextually accurate LLM drafts. Users reported that the frequent inaccuracies of the LLM drafts gave them concerns about the reliability of the LLM drafts, with some expressing that the current system could not be fully trusted. There was a suggestion that the LLM should refrain from generating an LLM draft if it was unsure, rather than providing potentially misleading information.

**2. User Experience and Feedback**

Some users appreciated the concept and potential of the LLM drafts, finding it useful as a starting point or for generating initial insights (e.g., summarizing inquiries). However, they acknowledged the LLM drafts still needed significant improvement as they did not necessarily solve the problems posed in the inquiries. Users suggested having more direct involvement in the improvement of the LLM drafts, such as the ability for users to mark correct and incorrect drafted replies, as feedback to the LLM and the developers. Additional training and system refinement were suggested to improve the overall effectiveness of the system.

## 4.3. Text similarity analysis

### 4.3.1. Semantic embedding similarity (SES)

We calculated the Semantic Embedding Similarity (SES) score for each case by converting the LLM draft and the sent reply into semantic embedding vectors and then computing the cosine similarity between them. A score closer to 1 indicates higher semantic similarity, while a score closer to 0 indicates semantic irrelevance. Figure 8 shows the distribution of SES scores obtained from the final dataset of 2,306 cases. Panels A and B display the results from two text-to-embedding models: text-embedding-ada-002 (ADA) and text-embedding-3-large (LARGE), respectively. Each distribution has been adjusted to its valid score range.

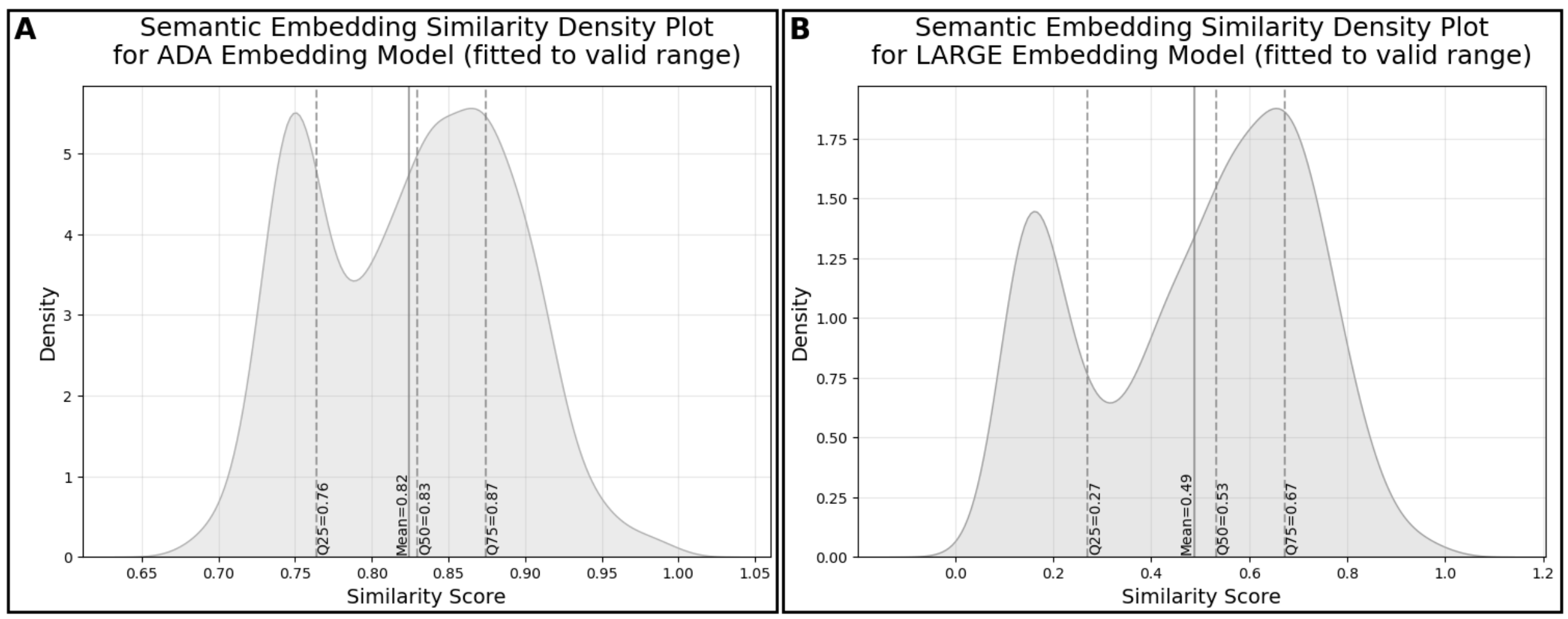

Figure 8. Cosine similarity score distributions (KDE) between LLM-generated text and case-handler-generated text embeddings using the ‘text-embedding-ada-002’ (ADA) and ‘text-embedding-3-large’ (LARGE) models

Most notably, both panels exhibit a bimodal distribution pattern. The distribution revealed two distinct groups: one with higher similarity scores where LLM drafts showed moderate alignment with sent replies, and another with lower similarity scores where LLM drafts and sent replies diverged significantly. From a utility perspective, the higher similarity group suggested that LLM drafts captured the semantic content of the sent replies to some extent, allowing case handlers to utilize them with moderate modifications. Conversely, the lower similarity group indicated that LLM drafts and sent replies were semantically divergent, requiring case handlers to undertake extensive rewriting. This polarized distribution suggested that the LLM's performance was inconsistent, being effective in some cases but inadequate in others.

A manual inspection of cases with low similarity scores revealed that they often included generic content not directly relevant to the case, such as requests for more context, clarifications of the question, or suggestions to contact the relevant office or expert. Such generic text was dissimilar to the specific, contextualized answers required by the inquiries, which resulted in very low similarity scores.

The two models showed distinct characteristics. The ADA model (Panel A) produced a compressed distribution of scores mainly ranging from 0.65 to 1, whereas the LARGE model (Panel B) covered the full 0-to-1 range. Given its wider dynamic range and thus greater capacity for nuanced detection of semantic differences and similarities, we selected the LARGE model for our main analysis. To ensure the robustness of our results, we repeated the text-to-embedding transformation with the LARGE model. The difference between these two embedding sets was negligible ($1.32 \times 10^{-6}$), which supported the reliability and reproducibility of the SES scores.

The descriptive statistics of semantic embedding similarity scores in Table 8 provide additional insights into these findings. The standard deviation (0.2260) highlighted significant variability in text similarity, reflecting a wide range of LLM's performance across cases. This variability corresponded to the observed bimodal distribution. The bottom 25% of cases had low similarity scores below 0.2704, suggesting that drafts and replies differed significantly. The top 25% of cases showed similarity scores above 0.6717, indicating that the LLM performed relatively well in these cases. With a median of 0.5331 and the top 25% boundary at 0.6717, approximately one-quarter of all cases were concentrated between scores of 0.53 and 0.67. This suggested that a substantial number of cases exhibited moderate similarity between LLM drafts and sent replies. The mean of 0.4879 being slightly lower than the median of 0.5331 reflected the influence of cases with considerably low scores.

This score distribution suggested that LLM drafts have not yet reached a level where case handlers can use them without significant modifications. This reflected the high standards of accuracy and domain expertise required in maritime industry communications.

### 4.3.2. LLM as a judge (LAAJ)

We had the LLM evaluate the similarity between LLM drafts and sent replies across three criteria: overall similarity, language similarity, and interpretative similarity, and also calculated a composite similarity score by averaging these results. The evaluation was conducted on a 1-7 Likert scale, where 1 indicated low similarity and 7 indicated high similarity. Figure 9 shows the LAAJ score histograms for these evaluation criteria. Since the scores represented averages from three independent runs for each case, they appeared as discrete distributions with some variation.

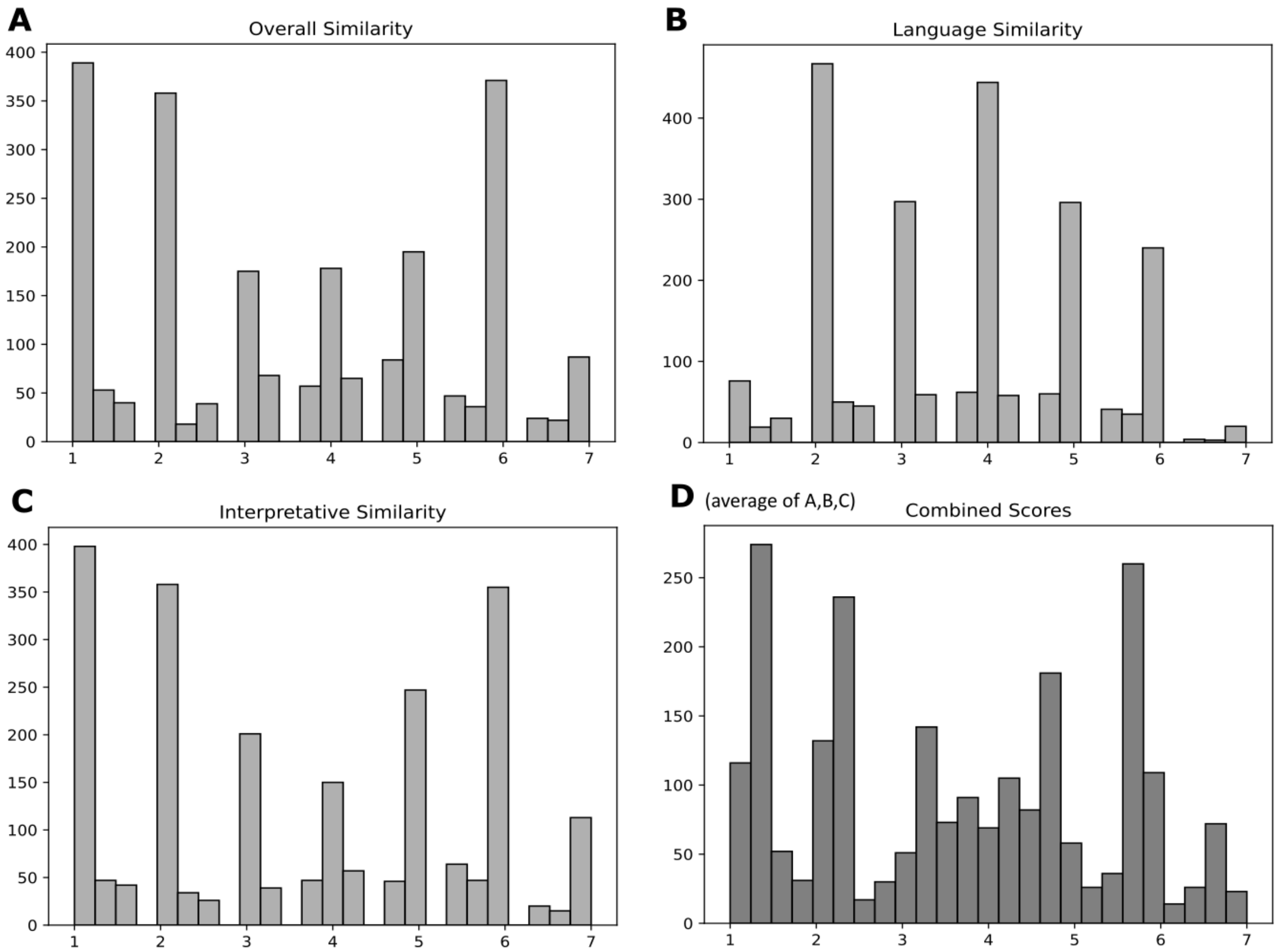

Figure 9. LAAJ histograms for overall similarity, language similarity, interpretative similarity and combined scores, respectively

Overall similarity exhibited, in general, a bimodal distribution with high frequencies at scores 1-2 and at score 6. Interpretative similarity showed a generally similar pattern, also having high frequencies at scores 1-2 and at score 6. Language similarity displayed a somewhat different pattern, showing high frequencies at scores 2 and 4. The similar patterns between overall similarity and interpretative similarity suggested that overall similarity largely reflected semantic and interpretative aspects of the text. The relatively moderate levels in language similarity indicated that case handlers frequently modified specific expressions and language, while the bimodal distribution in interpretative similarity suggested that the LLM either properly grasped the core meaning of inquiries or failed to do so.

The composite score distribution also formed a generally bimodal pattern, showing multiple distinct peaks. The highest peak occurred at score 1 representing "Not similar at all," a second notable peak appeared at score 6 representing "High similarity," a third peak emerged at score 2 in the "Low similarity" range, while two smaller peaks appeared in the "Moderate similarity" range of scores 3-5. This distribution with major peaks at the lower and higher ends suggested that the LLM's performance was inconsistent, being effective in some cases but inadequate in others.

For comparison, a system demonstrating consistently higher similarity would show a markedly different pattern: a unimodal distribution heavily skewed toward higher similarity of scores 6-7, with minimal cases in the lower similarity ranges. The actual multi-modal distribution we observed, with significant peaks at both extremes, indicated room for improvement in the context of professional communication where accuracy and precision are crucial. Despite the presence of some highly similar responses, this pattern indicated the system had not yet achieved the consistent, reliable performance required in professional maritime communications.

To ensure the reliability of these similarity assessments, we conducted three independent evaluation runs, and the consistency of these results was statistically confirmed. ANOVA results showing no statistically significant differences between runs, very high Kendall's W values, and visually consistent score distributions (see Appendix F for detailed information) suggested that the LLM applied consistent evaluation criteria across all runs, supporting the reproducibility and reliability of LAAJ.

The descriptive statistics of the composite scores in Table 8 provided 'complementary insights into the LAAJ results. The LAAJ evaluation yielded a mean composite score of 3.63 and a median of 3.55 on the 1-7 scale, suggesting a moderate level of similarity between LLM drafts and case handlers' sent replies. While moderate similarity scores might appear encouraging, taking a conservative interpretation appropriate for safety-critical maritime industry communications, these scores indicated insufficient evidence to claim high textual similarity between LLM drafts and case handlers' sent replies.

### 4.3.3. Additional analysis (triangulation): Comparing similarity scores between SES and LAAJ

We sought to establish the reliability of our textual similarity assessments through methodological triangulation using SES and LAAJ. This additional triangulation analysis was to emphasize its role in validating the reliability of the textual similarity assessments, highlighting the complementary perspectives and strong correlation as evidence of robustness. While both methods indicated moderate average similarity levels between LLM drafts and case handlers' replies (SES mean = 0.4879; LAAJ mean = 3.63 on 1-7 scale), they offered complementary perspectives on the similarity patterns. The strong correlation between these methods (Pearson $r = 0.8128$, $p < 0.001$; Spearman $\rho = 0.8025$, $p < 0.001$) provided robust evidence that both methods were capturing related aspects of text similarity, despite their different methodological foundations.

The comparison of score distributions (Figure 10) revealed both commonalities and differences in how these methods characterized similarity patterns. The SES method produced a distinct bimodal distribution, clearly separating cases into higher and lower similarity groups. In contrast, the LAAJ method yielded a more nuanced multi-modal distribution, suggesting finer distinctions across similarity dimensions. This difference reflected the inherent characteristics of each method: SES provided a continuous measure of semantic similarity, while LAAJ offered discrete, multi-faceted evaluation capturing distinct aspects of text comparability.

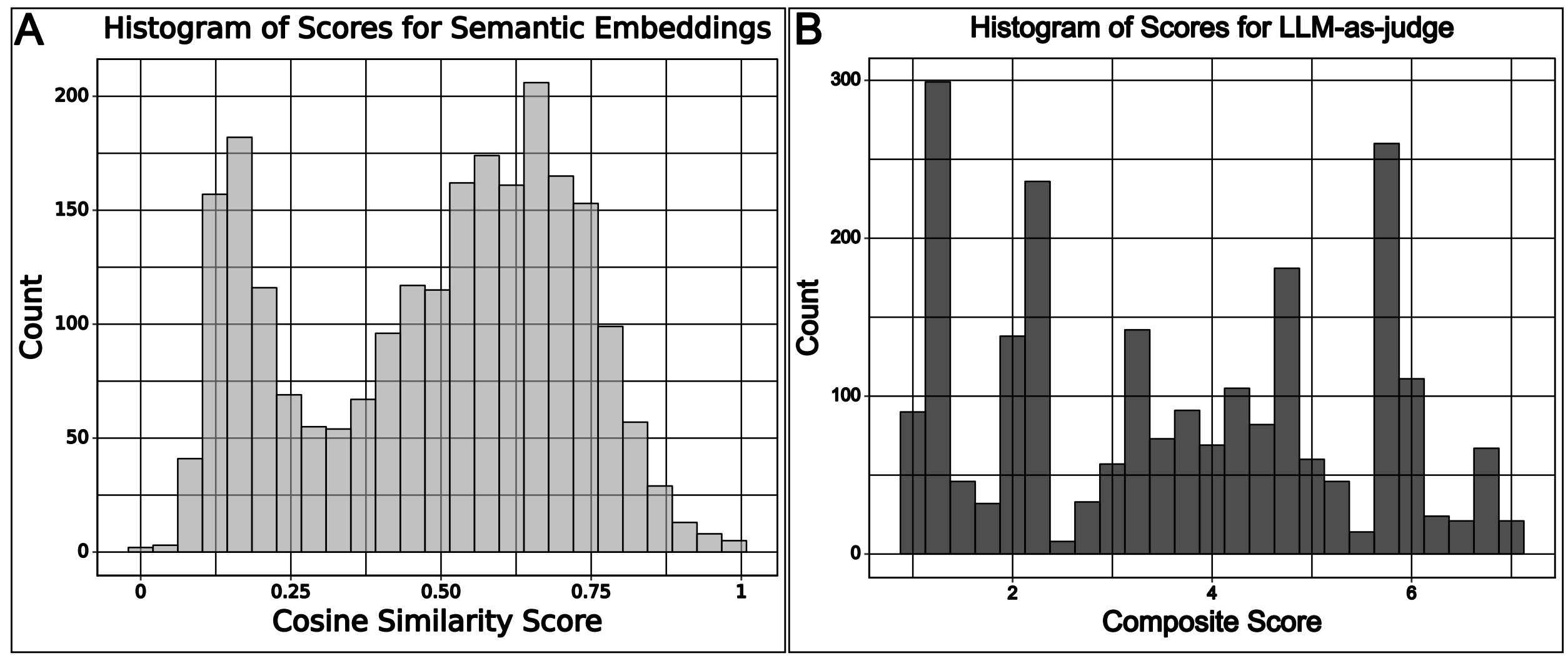

Figure 10. Histograms of SES and LAAJ scores

Table 8. Summary statistics for SES Score and LAAJ Composite Score

| | *Cosine Similarity Score (Semantic embedding)* | *Composite Score (LLM-as-judge)* |
|---|---|---|
| Mean | 0.4879 | 3.6307 |
| Median | 0.5331 | 3.5556 |
| Std Dev | 0.2260 | 1.7245 |
| Min | 0.0089 | 1.0000 |
| 25%: | 0.2704 | 2.0000 |
| 50%: | 0.5331 | 3.5556 |
| 75%: | 0.6717 | 5.0000 |
| Max | 0.9969 | 7.0000 |

To validate the reliability of this comparative analysis, we examined the statistical properties of both methods (Table 8). Both methods demonstrated considerable range in their measurements (SES: 0.0089-0.9969; LAAJ: 1-7), with similar patterns of central tendency and spread relative to their respective scales. The consistency in these patterns, combined with the strong correlations between methods, provided strong evidence for the robustness of our similarity analyses.

This triangulation of methods suggested that while SES and LAAJ should not be used interchangeably due to their distinct measurement characteristics, their strong correlation and complementary insights strengthened our overall analyses of text similarity between the LLM drafts and case handlers' sent replies. The convergence of results across these methods reinforced our conservative interpretation regarding the moderate levels of text similarity observed.

# 5. Discussion

Our real-world case study explores how case handlers, human experts in the maritime industry, perceive the usefulness of LLM-generated draft replies (i.e., LLM drafts) in responding to stakeholder inquiries. Our research questions were:

1. How do case handlers perceive the usefulness of LLM drafts, and how can these drafts support their workflows?
2. How similar are the texts generated by the LLM to the final replies sent by case handlers, and what does this similarity imply about the drafts' potential utility?

In this section, we discuss the key findings derived from a preliminary qualitative study, a survey study, and a text similarity analysis, while also identifying potential directions for future work.

It is important to note that the LLM draft function had been operational for only a few months at the time of our study. While this might be considered premature especially for assessing long-term impacts, the goal of our research was to establish a baseline. This baseline serves not only as a reference point to evaluate future advancements in technology and user perceptions but also as a foundation for tailoring improvements to the LLM drafts.

By analyzing our findings, we aim to start addressing the broader challenge of bridging the gap between user expectations and the current LLM functionality in general, and for the DATE system specifically. The findings can inform targeted improvements to the DATE system, helping it meet the needs of case handlers more effectively.

This approach emphasizes using the findings proactively, not just as benchmarks but as actionable guidance for developing the technology further, enhancing its relevance, and addressing areas where user expectations remain unmet.

## 5.1. Case handlers' perceptions of the LLM drafts

Our findings on perceptions of the LLM drafts can be grouped into three interrelated themes: perceived accuracy and relevance, attitudes toward adoption, and perceived impact on stakeholder trust. It is important to note that the survey only indicates how many case handlers had tried the LLM, not how frequently they used it, as the system was still in its early experimental phase at the time of the study. Moreover, the survey does not capture the perspectives of non-respondents or respondents who had not yet tested the drafts. Finally, prior research has shown that humans often hold biases against LLM-generated texts (Asscher and Glikson 2023; Harasta, Novotná, and Savelka 2024; Yihe Liu et al. 2022), and such biases should be considered when interpreting our findings.

### 5.1.1. Perceived accuracy and relevance

In general, the preliminary study and survey findings suggest that case handlers who tested the experimental LLM draft functionality feel that it requires significant improvement in terms of accuracy and relevance. Survey results indicate that case handlers still rely heavily on their own expertise or that of their colleagues, particularly when dealing with uncertainties (Figure 6). This reliance is likely influenced by the specialized nature of the maritime industry and the high level of expertise required for case handling. The complexity of stakeholder inquiries in this domain often necessitates niche expertise or a combination of specialized knowledge to provide accurate responses, likely called as tacit knowledge (Sporsem et al. 2023).

This raises important questions about how an LLM can be effectively utilized to handle inquiries of such complexity (Hodne et al. 2024; Reddy, Sinha, and Park 2024; Ulfsnes, Mikalsen, and Barbala 2024). In the survey findings, case handlers provided several suggestions for improving the LLM drafts. These improvements include broadening the system's data sources beyond historical cases to incorporate additional relevant materials – such as attachments – and integrating user feedback at the case level. For example, case handlers can mark text as correct or incorrect, enabling the system to refine its future LLM drafts. This feedback can then be used to improve the system's ability to retrieve and present relevant information (G. Gao et al. 2024; Shankar et al. 2024), ultimately aiming for its enhanced accuracy and relevance across a broader range of inquiries.

Despite these challenges, the survey findings also highlight several positive perceptions of the LLM drafts. Case handlers rated the language quality of the drafts as high and noted that the drafts helped them work more efficiently and maintain consistency in their replies to the stakeholders. These findings indicate that, while accuracy and relevance remain areas for improvement, users already perceive other aspects of the LLM drafts, such as linguistic quality and time-saving potential, as valuable.

### 5.1.2. Skeptical but curious

Our survey findings are in line with our preliminary findings, especially in identifying the attitudes of case handlers as "skeptical but curious". This skepticism towards the LLM drafts is an important and healthy safeguard against overreliance while encouraging the advancement of the LLM's utilization to create more accurate and relevant LLM drafts (Duenas and Ruiz 2024; Zhai, Wibowo, and Li 2024). This openness to innovation is likely related to the fact that the LLM drafts are developed in-house and the case handlers are involved and encouraged to explore the new function critically. Importantly, senior managers are supportive and engaged in the development of the LLM drafts. Such support is crucial; managerial endorsement plays a significant indirect role by building trust among employees, which in turn increases use of the AI-enabled systems (Korzyński et al. 2024). It is worth mentioning that the deployment of the LLM drafts follows an earlier implementation of two natural language processing (NLP) solutions based on Term Frequency-Inverse Document Frequency (TF-IDF) on the DATE systems, which had been well-received and perceived as effective. In addition, the LLM drafts' deployment has been communicated as a collaborative and exploratory effort, emphasizing that it is an initial step rather than a finalized product. The LLM drafts have not been overstated as a revolutionary AI-enabled system; rather, both the design and communication strategies have been deliberately cautious, acknowledging potential limitations and framing the deployment as a learning process. Such strategies are followed by the iterative design process, incorporating improvements based on real-world use and feedback. This strategies and framing may influence the creation of a more favorable environment for deploying the new system (i.e., the LLM drafts).

Our case study also highlights a broader organizational motivation for piloting the LLM functions: the recognition that AI-enabled systems are becoming ubiquitous and cannot be ignored. By actively engaging with these technologies, organizations can develop internal expertise, better understand the associated risks, and identify opportunities to harness AI-enabled systems effectively (Sarri and Sjölund 2024). This proactive approach positions organizations to embrace AI-enabled systems responsibly while mitigating potential drawbacks.

Having said that, the fact that only 74 of 198 case handlers responded to the survey and only 51 case handlers had tried the LLM drafts at the time of the survey can mean that those case handlers are the early adopters (a.k.a. champions). Early adopters are likely to have more optimistic attitudes towards the technology (Haque et al. 2022). This is a limitation in our survey and the findings should be read in this context. If the participants are indeed skewed towards more optimistic perceptions, the actual reception of the technology among the broader case handler base may be less favorable than the data suggests. We will need further investigation into barriers or hesitations among case handlers who did not respond and the non-user case handlers. Future work can also explore factors that influenced the decision whether to use the LLM draft. This can provide valuable feedback on the technology itself and help us understand the socio-technical factors surrounding the perceived usefulness of the LLM drafts.

### 5.1.3. Stakeholders' trust

The only statistically significant difference we found in the responses is between the case handlers who believe that the new function of LLM drafts will decrease the stakeholders' trust in the organization (the "decrease" group) and those who believe it will increase (the "increase" group) or have no impact (the "no impact" group). The "decrease" group having the lowest mean scores shows that these case handlers perceive the LLM drafts as less aligned with their expectations or less capable of accurately addressing stakeholder inquiries. This perception likely reflects concerns about the quality or appropriateness of the LLM drafts, which may lead to fear of misunderstandings or dissatisfaction among stakeholders ultimately impacting trust in the organization. In contrast, the "increase" and "no impact" groups likely see the LLM drafts as being sufficient to support or maintain stakeholder trust. This difference in perception suggests that some case handlers have more confidence in the LLM than others, highlighting the need to address concerns about the quality and reliability of the LLM drafts. Addressing case handlers' specific concerns can contribute to building trust within the organization and with stakeholders. Accordingly, future work can explore underlying issues in the different perceptions among case handlers to tailor improvement efforts.

## 5.2. Similarity between LLM drafts and sent replies

### 5.2.1. Variability in text similarity

Text similarity analyses between LLM drafts and case handlers' sent replies provide valuable insights into the LLM's performance. The similarity scores exhibit a bimodal distribution suggesting variability of similarity levels. This distribution highlights two primary patterns in case handlers' use of LLM drafts: cases where the drafts are largely retained and cases where extensive revisions are required. Specifically, the group of highly similar texts suggests that the LLM performs well, enabling case handlers to use the LLM drafts without making substantial revisions. Conversely, the group with lower similarity may indicate instances where the LLM lacks sufficient contextual understanding or information, requiring case handlers to significantly adapt the LLM drafts to meet the inquiry's specific needs.

However, the survey findings appear more weighted towards the lower similarity group and much less towards the higher similarity group. The survey suggests that most LLM drafts require significant changes (Table 6). The respondents who only modified the LLM drafts slightly or none at all, may correspond to the high-similarity group. The widespread perception of inaccuracy may help explain why extensive modifications are frequently needed, as the survey findings reveal that the majority of case handlers (86.28%) disagree or strongly disagree with the statement that the LLM drafts are accurate (Table 6).

Importantly, the text similarity findings reveal significant variability in the similarity between LLM drafts and case handlers' sent replies. This suggests that the LLM's performance is inconsistent, excellent in some cases but inadequate in others. This is particularly relevant in the complex and safety-critical maritime industry, which demands high levels of accuracy and precision in the communication of critical information. Therefore, we find that these LLM drafts are not yet mature enough for use in high-risk industries without human experts.

The variability in similarity between the LLM drafts and the case handlers' sent replies may be influenced by several factors. One key limitation is that the LLM system cannot process attachments, which may contain critical information necessary to address the inquiries. Additionally, the LLM lacks access to vital resources available to case handlers such as internal reports, relational knowledge (Ulfsnes, Mikalsen, and Barbala 2024), historical context, and access to other experts. Outdated past cases – due to updated or new regulations, for example – also require case handlers to amend the LLM drafts to ensure compliance and accuracy. In these cases, case handlers leverage their more comprehensive understanding of the inquiries to refine the drafts. This directly impacts the text similarity results. In addition, variations in writing and communication styles between the LLM and individual case handlers may further contribute to the observed variability.

Although our case study does not specifically investigate a potential link between similarity levels and the complexity of inquiries, preliminary findings and survey results suggest that inquiry complexity plays a significant role. Considering the inherently complex nature of the maritime industry, it is plausible that higher similarity between LLM drafts and sent replies corresponds to less complex inquiries, for example, inquiries related to asking for confirmations that certain changes to be made in the documents. This aligns with the notion that simpler, more straightforward inquiries may require fewer adjustments by case handlers, while more multifaceted cases may lead to more adjustments. Accordingly, differences in handling straightforward versus complex inquiries may be another factor influencing the variability in the similarity.

### 5.2.2. Pathways for improvement

The wide variability in similarity scores highlights several areas for potential improvement in the system. For example, the tendency of the LLM to produce generic or insufficient responses, as also reported in the preliminary and survey findings, can be mitigated by improving the model's contextual understanding. The inability of the current LLM system to handle attachments or multi-turn conversations that limits its applicability to more nuanced inquiries may be addressed through improved data integration and dialogue management. The improved data integration may also address the limitation that the LLM does not have access to the same resources as case handlers. In addition, removing outdated cases may improve the accuracy and relevance of the LLM drafts.

Looking ahead, several technical advancements may improve the consistency of the LLM draft responses to inquiries with different levels of complexity. For example, computation enhanced generation could integrate external computational tools and models with LLMs, enhancing their accuracy in tasks that require precise calculations or domain expertise (Paranjape et al. 2023). The expanded context window capabilities seen in newer models, including those that can handle up to 1-2 million tokens, could potentially reduce reliance on retrieval for handling complex inquiries. Multimodal capabilities would enable processing of various data types often present in maritime documentation, enhancing the system's ability to comprehend and respond to diverse information formats (Yin et al. 2024). The development of smaller, more efficient models that maintain performance – through techniques like distillation and quantization – could make the system more practical and cost-effective (Shridhar, Stolfo, and Sachan 2022). Dynamic prompting that creates prompts in real-time based on user behavior and context could improve the system's adaptability and responsiveness (Yang et al. 2023). Finally, better integration with operational workflows through LLMOps could enhance the system's reliability and maintainability in production environments (Diaz-De-Arcaya et al. 2024).

## 5.3. Beyond generating accurate and relevant LLM drafts

The preliminary study and survey findings reveal that case handlers continue to place significant emphasis on the accuracy and relevance of LLM drafts when handling inquiries that are often highly specific and complex. Interestingly, despite the relatively negative perception of the LLM drafts' accuracy and relevance, the case handlers express a willingness to recommend the drafts to their colleagues, engage in frequent discussions about them, and intend to continue using them over the next 12 months (Table 6). These results suggest that the value of LLM drafts is likely to extend beyond merely generating accurate and relevant replies. For example, the preliminary findings indicate that case handlers use the drafts as a tool to generate ideas and explore potentially necessary information for crafting responses. This example highlights the potential of LLM drafts to serve as creative and exploratory aids in workflows (Ulfsnes, Mikalsen, and Barbala 2024). Below, we provide some known strengths and limitations of both the LLM drafts and case handlers, with the purpose of identifying potential use cases where the LLM drafts can provide value beyond generating accurate and relevant responses (Table 9).

Table 9. Identified key strengths and limitations of the LLM drafts and case handlers

| | Strengths | Limitations |
|---|---|---|
| **LLM drafts** | Produces high-quality, professional language in responses efficiently. | Limited to processing text inputs; cannot analyze other formats, such as images or attachments. |
| | Facilitates idea generation and brainstorming. | Relies exclusively on past cases as its data source, lacks access to external or real-time information. |
| | Efficiently handles inquiries involving straightforward instructions and readily available explicit knowledge. | |
| | Ensures consistency in responses and improves efficiency in drafting replies. | |
| | Quickly retrieves and incorporates relevant information from past similar cases. | |
| **Case handlers** | Can process and integrate multiple types of input, including text, images, and attachments. | Reliance on manual searches for past similar cases, which can be time-consuming. |

| | Access to other human experts for consultation and collaboration. | Variability in communication styles, leading to inconsistencies in responses. |
|---|---|---|
| | Possess relational knowledge about stakeholders, including familiarity with their preferences and how best to respond. | Risk of misinterpreting lengthy, complex inquiries. |
| | Utilize a wide range of information sources beyond the database, including institutional memory and industry expertise. | |

By exploring these examples, we can find new avenues for the application of LLM drafts and expand their utility beyond generating accurate replies (Duenas and Ruiz 2024). For example, the LLM drafts can synthesize insights from past cases, uncovering patterns or information that might otherwise go unnoticed. Additionally, the LLM can function as a knowledge hub, providing case handlers with quick access to precedents and facilitating deeper analysis of similar past cases. Another key advantage is the LLM's ability to reduce time spent on routine or preliminary tasks – such as reviewing lengthy and/or complex inquiries – so that the LLM drafts can assist in confirming or correcting case handlers' interpretations of such inquiries (Nasseri et al. 2023). By leveraging its capacity to process and summarize large volumes of text, case handlers can focus more effectively on decision-making and crafting tailored responses. This functionality could also support the training of more junior case handlers by offering templates or examples to guide their work.

In addition to improving workflows, the LLM drafts enhance the overall quality of written communication. Survey findings suggest that case handlers' perception of the language quality of LLM drafts is high, indicating their potential to set benchmarks for clarity and professionalism. By creating templates for common cases, LLM drafts can improve efficiency while also standardizing communication. This standardization ensures consistent, high-quality responses aligned with organizational standards, which can enhance stakeholder satisfaction and the organization's reputation.

Interacting with the LLM drafts also offers an opportunity for case handlers to become more familiar with the technology's capabilities and limitations. As this familiarity grows, case handlers may discover novel use cases where the LLM can provide even greater value, further integrating it into their workflows (Ulfsnes et al. 2024). For example, in a case study that aimed to identify the intention to use LLMs / generative tools by employees of an IT department in relation to their work found that employees who use the tools more frequently perceive the tools as more useful and thus have stronger intention to use the tools (Agossah et al. 2023).

## 6. Lessons learned, identified gaps and directions for future work

We have identified limitations in our case study, as well as lessons learned, gaps, and directions for future work. Although we validated the text similarity results within and across the LAAJ and SES, we acknowledge that a fully accurate evaluation remains difficult (Shankar et al. 2024). We did attempt a manual/human evaluation of a very small subset of cases from the text similarity analysis to calibrate its results (Virk, Devanbu, and Ahmed 2024), focusing on the inquiries, LLM drafts, and sent replies. However, this proved extremely challenging due to the nuanced content. In addition, scaling up such manual evaluation would prove even more challenging considering the volume of our dataset (N = 2,306). We concluded that completing a comprehensive evaluation for all cases was nearly impossible without extensive domain-specific expertise. This is because of the highly specialized nature of the inquiries and LLM drafts that necessitates specific domain expertise (Sporsem et al. 2023). The inquiries span over 650 maritime fields of expertise, reflecting the complexity and breadth of knowledge required to understand them. Although our dataset includes queries across these 650 categories, we did not analyze variance between categories, as this was beyond the scope of the study. The mention of these categories is intended solely to illustrate the diversity and complexity of the maritime domain. Our primary focus was instead on establishing baseline perceptions of the LLM drafts and examining overall similarity patterns. Accordingly, future work should explore more robust, mixed-methods evaluation approaches that incorporate domain expertise, case handlers' reasoning, and qualitative insights to assess LLM-generated content more comprehensively (Shankar et al. 2024). In relation to this, our focus on the maritime industry may limit the generalizability of the findings to other domains (Zou et al. 2023). Future research is needed to explore LLM utility across different industries and use cases beyond maritime inquiries. For example, conducting comparative studies in other highly specialized domains to understand how LLMs perform across diverse contexts and tasks to enable cross-domain learning (Zhou et al. 2024).

While similarity metrics can provide a useful starting point for evaluating the alignment between LLM drafts and sent replies, they may not be sufficient to fully capture the LLM's utility. For example, lower similarity texts could reflect cases where the LLM drafts served as a starting point for generating nuanced replies, thus demonstrating some value in utility even though the similarity scores were lower. Additionally, higher text similarity scores may not always correspond to high utility. Consider a scenario where a stakeholder inquires, *"What documents do I need to submit for a cargo claim?"* The LLM draft responds, *"What type of cargo is involved, which flag, is there* *damage or loss, which location, and under what circumstances?"* While this response aligns with the general process of gathering additional information and the case handler sends it with minimal edits, resulting in a higher similarity score, the inquiry remains incomplete. The response does not directly address the stakeholder's request for a document list and instead prompts further clarification, necessitating follow-up interactions to resolve the inquiry.

Another limitation with text similarity metrics is ambiguity in dimensions and extent of similarity (e.g., language, interpretability, or contextual similarity). LLM drafts fundamentally operate as language tools, and the nuances of language can lead to complexities that computational text similarity metrics often fail to capture. A phrase can have the same underlying meaning but be expressed in entirely different ways. Conversely, phrases that appear linguistically similar may carry significantly different meanings depending on context. As an example, the LLM draft can respond to an inquiry of "*What is the status of my shipment*?" with "*Your shipment is currently in transit and will arrive on Thursday*", which the case handler might amend to "*The shipment is on its way and should reach you by Thursday*". While the LLM drafts and sent reply convey similar information, they are phrased differently, potentially resulting in a lower similarity score despite their equivalence in meaning and utility. Similarly, there might be ambiguity in interpretation: an inquiry that says, "*Do I need to file additional forms*?" can be responded by the LLM draft as "*No additional forms are needed*," in which is amended by a case handler into "*No, you do not need to submit extra forms at this stage*". While the LLM draft and sent reply are highly similar in language, the case handler's reply adds the clarification "*at this stage*", which acknowledges that further steps might require additional forms, an important distinction from the LLM draft.

Future work should explore complementary metrics, such as uncertainty measures (Ye et al. 2024) or confidence levels (Virk, Devanbu, and Ahmed 2024), which can provide a more comprehensive understanding of the LLM system's effectiveness. Importantly, future work should include a thorough investigation to explore the specific degrees and aspects of textual similarity represented by both higher and lower similarity scores, ideally through human-as-a-judge, to calibrate the computational evaluation metrics in real-world settings (Jung, Brahman, and Choi 2024; Virk, Devanbu, and Ahmed 2024). This deeper understanding would enable more precise and meaningful evaluations of the LLM performance. Additionally, analyzing whether certain patterns emerge in the generated LLM drafts and examining the factors that influence varying levels of text similarity with domain experts could provide valuable insights for system improvements.

Our case study lacks detailed investigation into case handlers' needs and interaction patterns with the LLM system. Incorporating user feedback at the inquiry and case level could provide valuable insights into where the system excels or struggles (G. Gao et al. 2024; Shankar et al. 2024). For example, analyzing the types of questions posed and how users engage with the LLM could help identify scenarios where the system performs effectively – such as responding to straightforward, simple inquiries – and areas where it is less effective, particularly with more complex cases. Our findings imply that the LLM is less suited to handling these complex inquiries, but unpacking its performance in these contexts requires a more targeted approach.

Collecting and mapping user feedback at the case level, as suggested by case handlers in the survey findings, could highlight the LLM's strengths and weaknesses and also identify areas where human intervention is most and least critical (G. Gao et al. 2024). This would support more effective human-AI collaboration and guide future system improvements (Tita A Bach, Kristiansen, et al. 2024). Hence, future work should focus on incorporating detailed user feedback at the inquiry and case levels to better understand how case handlers interact with the LLM system and where it succeeds or struggles (G. Gao et al. 2024). This may involve categorizing cases by complexity to map the LLM's performance and identify patterns in scenarios where it performs effectively or falls short (Hu et al. 2023). In a broader sense, future work should also focus on shifting the emphasis from designing for automation to designing for enhancing human-AI collaboration (Amershi et al. 2019; Vats et al. 2024). This involves developing frameworks and methods that embed collaboration-first principles into algorithms and interfaces that align AI-enabled systems with real-world needs and fosters systems that augment humans rather than displacing them (Tita A Bach, Kristiansen, et al. 2024; Seeber et al. 2020; Shneiderman 2020).

Our case study does not directly or in detail investigate potential implications of the integration of the LLM drafts into case handlers' workflows for expertise development (Choudhury and Chaudhry 2024; Crowston, Bolici, and e del Lazio Meridionale), particularly concerning knowledge sharing and competence building among junior or less experienced and senior or more experienced case handlers (Kernan Freire et al. 2024). The survey findings indicate that most case handlers rely on their own knowledge or consult colleagues also when in doubt, suggesting that the LLM drafts *currently* augment rather than replace traditional knowledge-sharing practices. However, a minority of case handlers perceived the LLM drafts to have influenced how they consult colleagues. This signals potential shifts in collaboration dynamics as LLM technology evolves. It is worth noting here that the LLM draft function had been operational for only a few months at the time of our study. As the LLM technology advances, future work should look more detail into how the LLM drafts influence collaboration between and among case handlers and other experts in the organization, and the implications to knowledge-sharing and competence building.

# 7. Conclusion

LLMs represent a significant opportunity to transform workflows in specialized domains. Our study shows that while LLMs can streamline processes and support case handlers as human experts, they are not yet mature enough for independent use in safety-critical applications, and human oversight remains essential. LLMs can play a valuable role as an augmentation rather than a replacement of human expertise, and system improvements, such as broader data integration and iterative user feedback, can enhance performance. However, incorporating tacit knowledge gained through experience remains challenging, if not highly improbable, in real-world applications.

Our findings demonstrate that case handlers' expertise and experience will remain to be critical for ensuring replies to stakeholder inquiries are relevant, accurate, and aligned with the regulatory and operational context of each vessel. In the maritime industry, stakeholder inquiries often require tailored, highly specialized knowledge. While LLM drafts were helpful in many cases, they still exhibited limitations, highlighting the indispensable role of human expertise for precision, contextual understanding, and relational knowledge.

We conclude that LLMs should augment, not replace, human decision-making. Final responsibility must remain with case handlers as the human experts. By leveraging LLMs thoughtfully and fostering human-AI collaboration, organizations can enhance efficiency while maintaining high standards of quality, accuracy, and relevance tailored to each case.

# 8. Acknowledgments

Authors would like to thank Dr. Martin Høy for providing the data for text similarity analyses, Dr. Caryl de la Serna for conducting the R analyses and supporting the interpretation of the results, the DATE system's team and management, and the participants. This manuscript involved the use of ChatGPT 4o and Claude 3.5 Sonnet to assist in generating ideas, improving language clarity, and translating to English. All content was rigorously reviewed and refined by the authors, with additional proofreading performed by a professional proofreader to ensure readability, accuracy, and quality. The pre-print of this manuscript is available on arXiv (Tita Alissa Bach, Babic, et al. 2024).

## 9. Funder information

This project is partially funded by the Norwegian Research Council (grant number: 309631).

## 10. Declaration of interest statement

Authors declare no conflict of interest.

## 11. Data availability statement

Raw or minimally processed data are not available due to proprietary, ethical, legal, and confidentiality restrictions, and due to terms of participant consent. Specifically, the following constraints apply:

- Proprietary and security data: The text similarity data used for analysis includes sensitive, proprietary information related to DNV customer inquiries and internal company data, which cannot be released for legal, privacy and security reasons.
- Ethical, confidentiality, and privacy concerns: The survey responses and qualitative data (interview transcripts and observation notes) cannot be shared publicly. Even after anonymization, the unique nature of the data, combined with potential cross-referencing ("slicing") of responses, carries a significant risk of re-identifying participants, thereby compromising the guaranteed anonymity and confidentiality of the DNV employees.

## 12. Authors contributions

TAB, AB, NP conceptualized the study, designed the research methodology, conducted analyses, and led the writing of the manuscript. TSp and RU conducted the preliminary method, contributed to the overall data analysis, and critically reviewed the manuscript. HSM as solution designer and TSk as process responsible provided technical expertise on DATE systems and case handlers, contributed to the interpretation of results, and reviewed the manuscript critically. All authors read and approved the final version of the manuscript.

## 13. Biographical note

TITA A. BACH received a PhD in behavioral and social sciences from the University of Groningen, the Netherlands, in 2012. She received a diploma in Top Tech Executive Education from Haas School of Business, University of California, Berkeley, CA, USA, in 2017. In January 2023, she joined Digital Transformation Team, Digital Assurance, Group Research and Development, DNV, Høvik, Norway, as a Principal Researcher. From 2011 to 2022, she was a Principal Researcher in the Healthcare program at the same organization. Her research focuses on several areas including human behaviors with technology in various environments, safety culture and cybersecurity culture, trust in technology and AI, AI deployment and adoption, human-AI interaction, and responsible AI.

ALEKSANDAR BABIC received a MS degree in electrical engineering and a MS degree in computer science from the University of Belgrade, Serbia, in 1999 and 2012, respectively, and a PhD degree in biomedical engineering from the University of Oslo, Norway, in 2019. He is currently a Principal Researcher with the Healthcare Program, Group Research and Development, DNV, Høvik, Norway. He has over two decades of industrial research and development experience and has worked on various projects, such as machine vision for 3D cameras and image fusion and deep learning for the applications of cardiac ultrasound. His research interests include the challenges associated with the implementation of AI in healthcare, including developmental and regulatory aspects, trustworthy and explainable AI, data quality and access, privacy, and algorithm development and robustness.

NARAE PARK is a researcher with an interdisciplinary academic background, holding MSc degrees in Language Technology from the University of Oslo and Computer Science from UiT The Arctic University of Norway, and an MA in Aesthetics from Seoul National University. She previously worked in the Digital Transformation Team, Digital Assurance, Group Research and Development at DNV, Høvik, Norway. Her research interests include language models, generative AI, extended reality, and the societal impact of these technologies.

TOR SPORSEM is pursuing a PhD in Computer Science at the Norwegian University of Science and Technology (NTNU), Trondheim, Norway. Since 2019, he has been employed as a Research Scientist at SINTEF in Trondheim. He specializes in sociological research methods, conducting qualitative studies through interviews and observations. His research focuses on the interaction between software developers and users, particularly the feedback loop during development, requirements engineering, and the creation of software incorporating AI components.

RASMUS ULFSNES is a research scientist at SINTEF Digital and a PhD candidate at the Norwegian University of Science and Technology (NTNU). His research focuses on the intersection of organizations, knowledge work, and artificial intelligence, with a particular emphasis on how organizations implement and adopt AI for experts and knowledge workers. Rasmus aims to understand the more systemic implications of the use of AI in organizations, through both qualitative methods, and digital data. He has previous experience from industry as both a Security manager and an Enterprise Architect.

HENRIK SMITH-MEYER received an MSc in Computer Science and Knowledge Systems from Stanford University in 1992 and an MSc from NTNU in 1989, building on a BSc from the University of Utah. In 2000, Henrik completed the Program of Strategic Leadership at the Norwegian Business Institute. With extensive AI experience, Henrik has implemented four Industrial AI production systems (in Hydro, Telenor, and Statnett) and contributed to the first Norwegian commercial AI company, Computas, managing areas such as Methods and Tools and multiple project deliveries. Worked on a tool framework recognized in Gartner's Magic Quadrant for AI-supported BPM and Enterprise Architecture, contributing patentable innovations. Henrik contributed to three AI research projects for the European Space Agency (ESTEC), and several for NRC and in EU Programs, advancing knowledge activation, interoperability, model-based design, HCI, and perceived value. Key contributor to team awarded DNDs prize of Business Intelligence 2018 for ML dev platform and first NLP solutions in DNV and DATE 2017. Solution Architect/Designer for DATE since 2013, in effect acting as production tool manager.

TORKEL SKEIE holds a bachelor's degree with Honours in Offshore and Mechanical Engineering from Heriot-Watt University, Scotland (1996). Since joining DNV Maritime in 1998, Torkel has worked extensively across various roles, including drawing approvals, quality assurance for survey and reporting data, and as a surveyor for new buildings, materials and components certification, and ships in operation. Since 2008, Torkel has been working as a technical expert, advising surveyors and customers on ship specific questions as well as being lead trainer for a course on ship classification. Since 2013, Torkel has worked as a team with Henrik to further develop the DATE system, focusing on user experience for both case handlers and stakeholders. Their work emphasizes intuitive design and agile development to deliver practical solutions.

## 14.References

# 15.List of tables, figures, and appendices

List of figures

List of appendices

# 16. Appendices

## Appendix A – The number of items exchanged per case

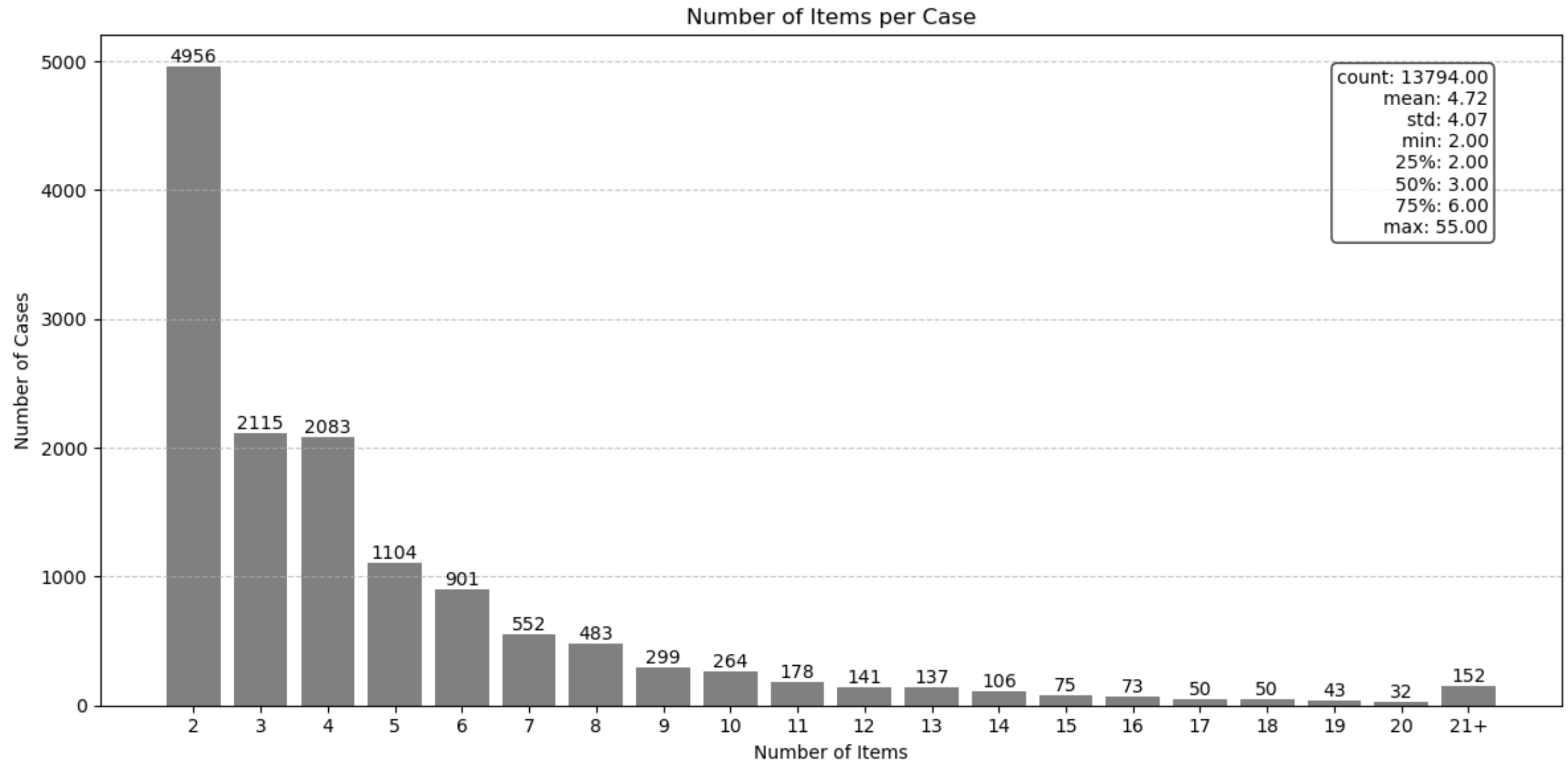

* Items represent individual communications within a case thread, categorized by content type (e.g., Question, Reply, InternalComment)

## Appendix B – The development of the questionnaire

To assess case handlers' experiences as users with the LLM drafts, a questionnaire was developed. The instrument was informed by in-depth discussions among the research team, DATE developers, and several case handlers to identify key aspects. In addition, several questionnaire items were adapted from established scales, including those by Brooke (1996), Davis (1989), and Venkatesh et al. (2003). This approach was to ensure that the questionnaire captured a comprehensive and relevant range of user perceptions and experiences while building upon existing measures. The questionnaire was tested on a few case handlers and iterated until all authors approved the final version.

The questionnaire consisted of 21 questions, in which 7 were categorized as demographic questions and 14 were measuring case handlers' experiences and opinions with the LLM drafts. The last demographic question measured how long a participant had been using the LLM drafts (i.e., a few days, a few weeks, a few months, or never). The survey was ended for those who responded with "never used the LLM drafts" after asking them to provide a reason (i.e., unavailable to them, chose not to use them, or other reasons followed by a free text comment field). Participants who had used the LLM drafts received the 14 questions (see Table). For these participants, the questionnaire was supposed to take 3-5 minutes to complete. The questionnaire was designed this way to minimize the time taken from case handlers' chargeable work hours.

**Demographic questions:**

1. Years of experience being a case handler
    - > 1 year
    - 1-5 years
    - 5-10 years

- 10-15 years
- >15 years

2. Office location
    - USA
    - Norway
    - Singapore
    - Greece
    - Germany
    - Poland
    - Prefer not to say
3. My native language is: [free text]
4. Age group
    - 20-30 years old
    - 31-40 years old
    - 41-50 years old
    - >50 years old
5. Gender
    - Female
    - Male
    - Prefer not to say
6. Current role
    - Engineer
    - Surveyor
    - Support
    - Consultant
    - Specialist
    - Management
    - Other: [free text]
7. How long have you been using the LLM drafts?
    - A few days
    - A few weeks
    - A few months
    - Never[2]: (followed by a branching question)
        - Because:
            - The LLM drafts are not available to me
            - The LLM drafts are available to me, but I choose not to use them.
            - Other reasons: [free text]

**Questions for those who responded with “a few days”, “a few weeks”, or “a few months”**

1. The LLM drafts are accurate.
    - Strongly disagree
    - Disagree
    - Agree
    - Strongly agree

[2] If a survey participant selects this, the survey stops here for them after they respond to the branching question.

2. I would recommend the LLM drafts to other case handlers.
    - Strongly disagree
    - Disagree
    - Agree
    - Strongly agree
3. The language in the LLM drafts is of high quality.
    - Strongly disagree
    - Disagree
    - Agree
    - Strongly agree
4. I find the summary of the LLM drafts useful.
    - Strongly disagree
    - Disagree
    - Agree
    - Strongly agree
5. My colleagues and I discuss the use of the LLM drafts frequently
    - Strongly disagree
    - Disagree
    - Agree
    - Strongly agree
6. I tend to modify the LLM drafts:
    - Extensively
    - Moderately
    - Slightly
    - None at all
7. I believe that the LLM drafts will:
    - increase the stakeholders' trust in the organization
    - decrease the stakeholders' trust in the organization
    - have no impact on the stakeholders' trust in the organization
8. Using the LLM drafts increases my productivity.
    - Strongly disagree
    - Disagree
    - Agree
    - Strongly agree
9. I intent to continue using the LLM drafts in the next 12 months.
    - Strongly disagree
    - Disagree
    - Agree
    - Strongly agree
10. Since I started using the LLM drafts, I consult with my colleagues:
    - less often
    - more often
    - about the same
11. I use the LLM drafts frequently.
    - Strongly disagree
    - Disagree
    - Agree
    - Strongly agree

12. Using the LLM drafts improves my responses to stakeholders in terms of (*multiple responses possible)*:
    - clarity
    - accuracy
    - efficiency
    - consistency
    - conciseness
    - stakeholder satisfaction
    - professionalism
    - other [free text]
13. When in doubt, I rely on *(multiple responses possible):*
    - the LLM drafts
    - my knowledge
    - my colleagues
    - other: [free text]
14. General comments or suggestions about the LLM drafts: [free text]

# Appendix C – The number of LLM draft generation attempts per case

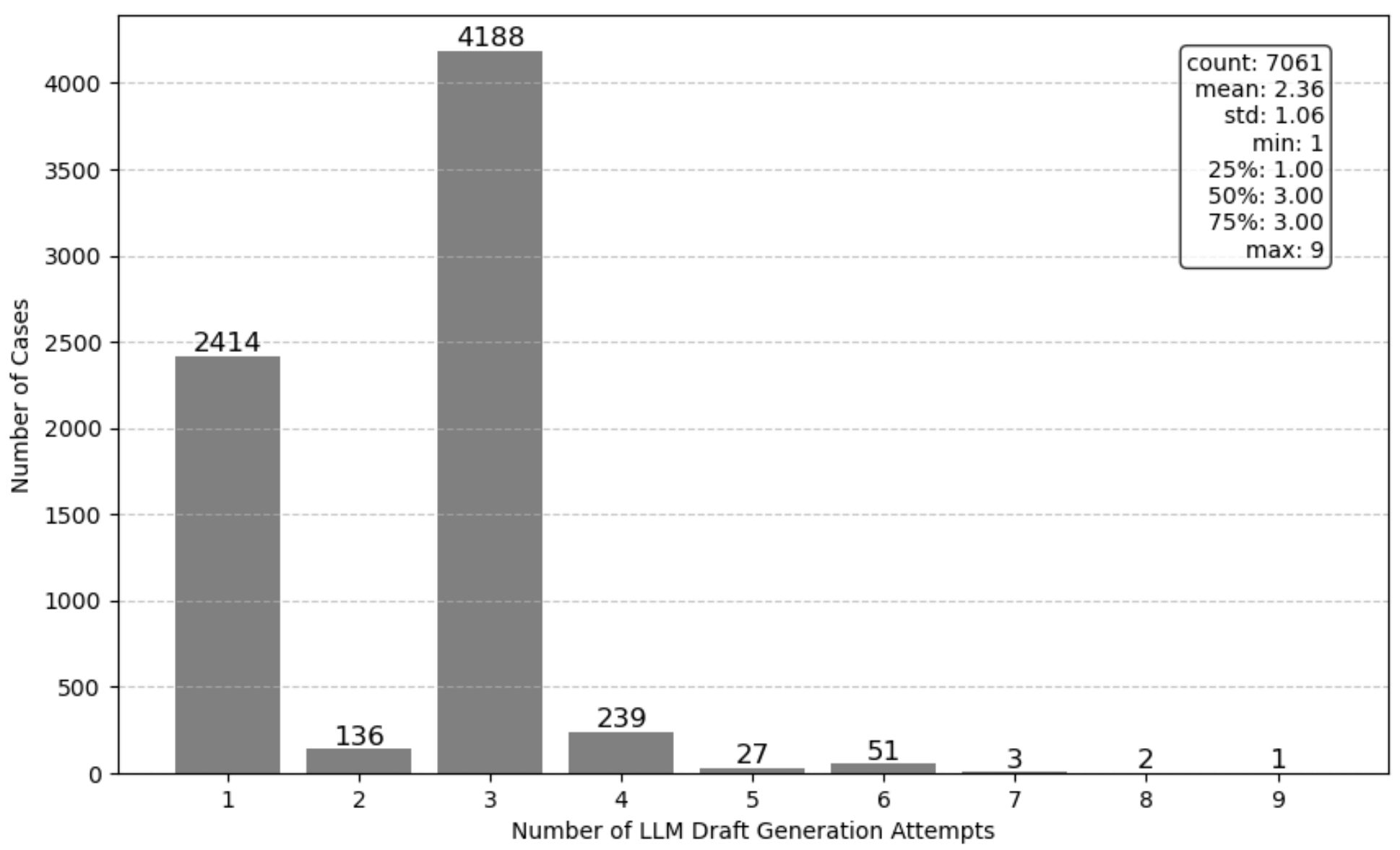

# Appendix D – LLM-as-judge (LAAJ) rating prompt

You are an AI assistant specialized in evaluating comparability of language and similarity of interpretation.

Analyze two replies to a given question and its corresponding subject to rate the level of comparability in language and rate the level of similarity of interpretation between the replies.

Evaluate each dimension listed above.

Determine an overall similarity score (1-7).

Provide a comprehensive explanation justifying all scores.

Ensure that explanation is valid string for JSON field.

Verify and validate your analysis for accuracy and completeness.

The rates are on scale 1-7 where:

* Comparability of language:
 - Not at all comparable = score in [1, 2]
 - Moderately comparable = score in [3, 4, 5]
 - Extremely comparable = score in [6, 7]

* Similarity of interpretation:
 - Not at all similar = score in [1, 2]
 - Moderately similar = score in [3, 4, 5]
 - Extremely similar = score in [6, 7]

Respond with a JSON structure:

```
{
 "interrater_analysis": [
  {
   "comparability_of_language": "1-7",
   "similarity_of_interpretation": "1-7"
   "overall_similarity_score": "1-7",
   "explanation": "Detailed justification of all scores",
  }
 ]
}
```

Ensure all values are properly enclosed in double quotes and remove unnecessary spacing and newlines.

The explanation should cover all criteria, including specific examples from the replies.

The verification step is crucial for ensuring accuracy and reliability of the analysis.

# Appendix E – Correlation Coefficient between the nine questionnaire items

| | The LLM drafts are accurate | I would recommend the LLM drafts to other case handlers | The language in the LLM drafts is of high quality. | I find the summary of the LLM drafts useful | My colleagues and I discuss the use of the LLM drafts frequently | I tend to modify the LLM drafts:[+] | Using the LLM drafts increases my productivity | I intent to continue using the LLM drafts in the next 12 months | I use the LLM drafts frequently |
|---|---|---|---|---|---|---|---|---|---|
| 1. The LLM drafts are accurate | 1.00 | 0.42** | 0.18 | 0.55*** | 0.06 | 0.40** | 0.45*** | 0.49*** | 0.41** |
| 2. I would recommend the LLM drafts to other case handlers | | 1.00 | 0.29* | 0.61*** | 0.30* | 0.16 | 0.59*** | 0.72*** | 0.63*** |
| 3. The language in the LLM drafts is of high quality. | | | 1.00 | 0.15 | 0.03 | 0.09 | 0.07 | 0.25 | 0.19 |
| 4. I find the summary of the LLM drafts useful | | | | 1.00 | 0.01 | 0.12 | 0.59*** | 0.63*** | 0.39*** |
| 5. My colleagues and I discuss the use of the LLM drafts frequently | | | | | 1.00 | 0.09 | 0.22 | 0.32** | 0.47*** |
| 6. I tend to modify the LLM drafts[+] | | | | | | 1.00 | 0.02 | - 0.02 | 0.04 |
| 8. Using the LLM drafts increases my productivity | | | | | | | 1.00 | 0.65*** | 0.62*** |
| 9. I intent to continue using the LLM drafts in the next 12 months | | | | | | | | 1.00 | 0.62*** |
| 11. I use the LLM drafts frequently | | | | | | | | | 1.00 |

***Significant at $p < 0.001$, **Significant at $p < 0.01$, *Significant at $p < 0.05$

[+] Extensively was scored as Strongly Disagree, Moderately as Disagree, Slightly as Agree, and None at all as Strongly Agree

# Appendix F – Inter-run consistency of LAAJ evaluations

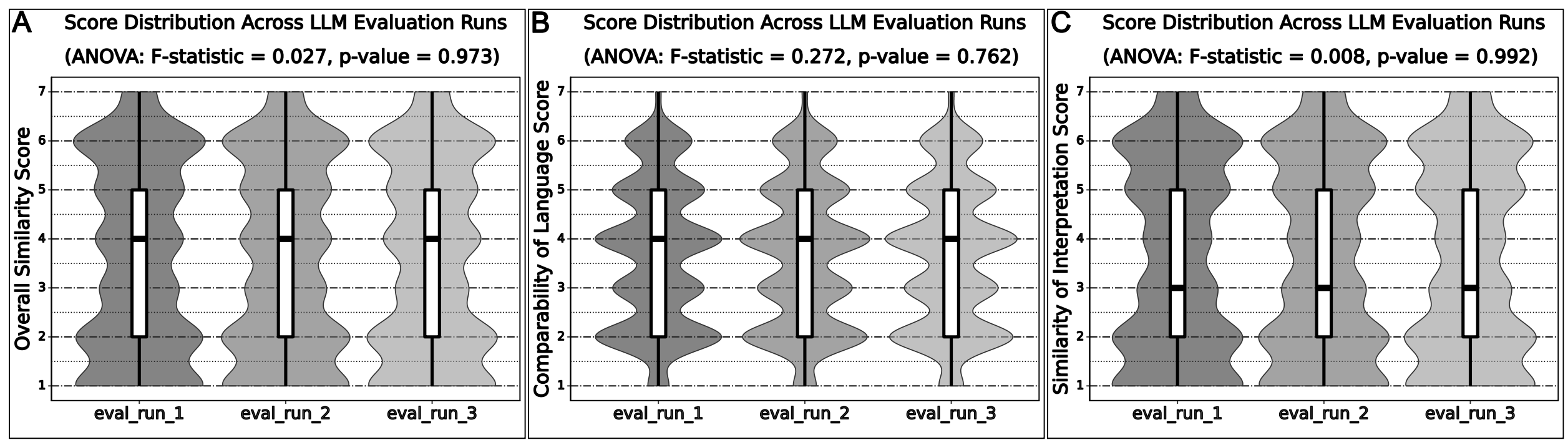

LAAJ score distributions (overall Similarity, comparability of Language, similarity of Interpretation) and ANOVA results across three LAAJ evaluation runs

| | Overall Similarity | Comparability of Language | Similarity of Interpretation |
|---|---|---|---|
| Kendall's W | 0.9527 | 0.9395 | 0.9543 |
| Chi-square statistic | 6587.9344 | 6496.9514 | 6599.1553 |
| Degrees of freedom | 2305 | 2305 | 2305 |
| P-value | 0.0000 | 0.0000 | 0.0000 |

LAAJ inter-run consistency statistics: Kendall's W, Chi-square statistics, p-values

Our analysis reveals high consistency across three dimensions of text similarity (overall similarity, comparability of language, and similarity of interpretation) across three evaluation runs. The violin plots illustrate consistent distributions across all three evaluation runs for each dimension. This visual consistency was corroborated by the ANOVA results which show no statistically significant differences between the runs. The high p-values strongly suggest that the differences observed between runs were likely due to random variation rather than systematic differences in the LLM's judgments.